\pdfoutput=1
\documentclass{jfm}
\usepackage{graphicx}
\usepackage{amsmath}
\usepackage{upgreek}
\usepackage{amssymb}
\usepackage{color}
\title{\pc{Effects of the finite particle size in turbulent wall-bounded flows of dense suspensions}}
\shortauthor{P.\ Costa, F.\ Picano, L.\ Brandt and W.-P.\ Breugem}
\shorttitle{Finite size effects in turbulent wall-bounded suspension transport}
\author{Pedro Costa\aff{1}
  \corresp{\email{p.simoes.costa@gmail.com}},
  Francesco Picano\aff{2},
  Luca Brandt\aff{3} \and
  Wim-Paul Breugem\aff{1}}

\affiliation{\aff{1}Process and Energy Dpt. -- Multiphase Systems, Delft University of Technology, Leeghwaterstraat 21, 2628CA, Delft, The Netherlands
\aff{2}Department of Industrial Engineering, University of Padova,\\ Via Venezia 1, 35131 Padua, Italy
\aff{3}Swedish e-Science Research Centre and Linn\'{e} FLOW Centre,\\ KTH Mechanics, SE-100 44 Stockholm, Sweden
}
\newcommand{\pc}[1]{#1}
\begin{document}
\maketitle
\begin{abstract}
We use interface-resolved numerical simulations to study finite-size effects in turbulent channel flow of neutrally-buoyant spheres. Two cases with particle sizes differing by a factor of 2, at the same solid volume fraction of 20\% and bulk Reynolds number are considered. These are complemented with two reference single-phase flows: the unladen case, and the flow of a Newtonian fluid with the effective suspension viscosity of the same mixture in the laminar regime. As recently highlighted in \cite{Costa-et-al-PRL-2016}, a particle-wall layer is responsible for deviations of the mesoscale-averaged statistics from what is observed in the continuum limit where the suspension is modeled as a Newtonian fluid with (higher) effective viscosity. Here we investigate in detail the fluid and particle dynamics inside this layer and in the bulk. In the particle-wall layer, the near wall inhomogeneity has an influence on the suspension micro-structure over a distance proportional to the particle size. In this layer, particles have  a significant \pc{(apparent)} slip velocity that is reflected in the distribution of wall shear stresses. This is characterized by  extreme events (both much higher and much lower than the mean). Based on these observations we provide a scaling for the particle-to-fluid \pc{apparent} slip velocity as a function of the flow parameters. We also extend the scaling laws in \cite{Costa-et-al-PRL-2016} to second-order Eulerian statistics in the homogeneous suspension region away from the wall. The results show that finite-size effects in the bulk of the channel become important for larger particles, while negligible for lower-order statistics and smaller particles. Finally, we study the particle dynamics along the wall-normal direction. Our results suggest that single-point dispersion is dominated by particle-turbulence (and not particle-particle) interactions, while differences in two-point dispersion and collisional dynamics are consistent with a picture of shear-driven interactions.
\end{abstract}

%
%\begin{keywords}
%dense turbulent suspensions, direct numerical simulation, finite size effects
%\end{keywords}
%
%
\section{Introduction}\label{sec:intro}
Turbulent transport of solid particles in a suspending Newtonian fluid is often found in natural and industrial contexts. Few of many well-known examples are sediment transport in rivers, sand storms, slurries, and the flocculation and sedimentation processes in the treatment of drinking water. In these examples and many others the flow can be dense, i.e., the solid volume fraction sufficiently high that short-range particle-particle interactions are dynamically significant. Moreover, the particles often have a finite size, i.e.\ comparable to or larger than the smallest relevant length scale of the flow.\par
The suspended particles may vary in size, shape, density, mechanical properties, etc. Moreover, the fluid can be non-Newtonian and the flow dynamics influenced by external fields (e.g.\ gravitational or magnetic), in addition to other specific characteristics of the different applications that can make the flow even more complex. Taking all these effect into account at once in a single configuration makes it difficult to disentangle the fundamental role of each parameter on the flow characteristics.
Even the simple \emph{canonical} case of the turbulent flow of mono-dispersed, rigid non-Brownian, neutrally-buoyant and spherical particles suspended in an incompressible Newtonian fluid is still subject of active research, with numerous fundamental questions remaining unanswered \citep{Prosperetti-JFM-2015}. The present work addresses this flow. In particular, we aim at describing finite-size effects on a dense turbulent suspension. Finite-size effects cause deviations of certain observables from the case where the dispersed phase can be modeled by an effective viscosity or by a more complex constitutive law for the stresses (i.e.\ for additives much smaller than the smallest relevant flow scale as colloidal particles, polymers and fibers).\par 
%
%From a modeling perspective, we require not only a realistic description of the effect of the suspending fluid on the particles (1-way coupling), but also the reciprocal effect of the particles in the fluid (2-way) and of long- and short-range inter-particle (and particle-wall) interactions (4-way) \cite{Elghobashi-ASR-1994}.
 Modeling the effects of the suspending particles in the flow through a localized source/sink of momentum \citep{Crowe-et-al-JFE-1977} is unrealistic when finite-size effects are important. In these cases the Navier-Stokes equations governing the suspending fluid should be coupled to the Newton-Euler equations governing the particle motion directly through the imposition of no-slip and no-penetration boundary conditions at the particle surface. This results in a relatively complex problem, difficult to tackle without three-dimensional and time-resolved flow data.\par
Throughout the years many studies on laminar shear flows laden with non-Brownian solid particles have been reported. In these cases one can take advantage of the linearity of the Stokes equations and achieve a rich variety of results by e.g.\ superposition of several canonical solutions. An iconic example is the effective viscosity of a suspension of non-Brownian spheres in the dilute and Stokes limit, derived by  \cite{Einstein1905} to be $\mu_e/\mu=1+(5/2)\Phi$, with $\mu$ being the viscosity of the suspending fluid, and $\Phi$ the bulk solid volume fraction. Many studies followed throughout the years \cite[see e.g.][for more rheological studies]{Stickel-and-Powell-ARFM-2005,Guazzelli-and-Morris-2011,Brown-et-al-RPP-2014}.\par
Most of the experimental works on particle suspensions have been limited to integral quantities, in particular the total wall shear. Often, the torque measured in a Taylor-Couette system required to keep a certain shear rate is used to measure effective viscosities and obtaining insight in the suspension rheology \cite[see e.g.][]{Bagnold-1954,Hunt-et-al-JFM-2002,Stickel-and-Powell-ARFM-2005}. Although many important results have been extracted from this approach, more detailed measurements of important features, such as the micro-scale organisation or the particle dynamics are challenging to obtain.\par
Lack of direct measurements at the micro-scale level gave room for important analytical studies that relate the bulk suspension behavior to the particle dynamics. A well-known example is the work of \citet{Leighton-and-Acrivos-JFM-1987}\pc{,} who introduced the concept of shear-induced migration due to irreversible inter-particle interactions\pc{,} to explain the migration of particles to the fluid reservoir in their Taylor-Couette experiments.\par
Great progress on the understanding of the flow dynamics at the particle scale has been achieved through numerical simulations. Here one can measure directly the suspension micro-structure and particle dynamics. In particular for flows in the Stokes regime, one can use very accurate and relatively inexpensive particle-based methods such as stokesian dynamics \citep{Brady-et-al-ARFM-1988} to understand in detail the bulk suspension behavior from a micro-scale perspective. Recent advances in experimental techniques made also possible direct measurements of a suspension micro-structure \cite[see e.g.][]{Blanc-et-al-JoR-2013}.\par
When inertial effects become significant, the governing equations for the fluid phase are non-linear and therefore analytical descriptions, numerical simulations and even experiments become more challenging. Moreover, if the Reynolds number is sufficiently high, the flow becomes turbulent, exhibiting chaotic and multi-scale dynamics. This makes the understanding of these flows even more difficult.\par
Despite this, significant progress has been made with regard to wall-bounded particle-laden flows in the turbulent regime, in the point-particle limit. Several studies have been carried out analytically \cite[e.g.][]{Maxey-and-Riley-PoF-1983,Reeks-JFM-1977}, numerically \citep{Soldati-and-Marchioli-IJMF-2009,Eaton-and-Fessler-IJMF-1994,Sardina-et-al-JFM-2012} and experimentally \citep{Fessler-et-al-Pof-1994}. As a result, a deep understanding of the mechanisms e.g.\ leading to preferential accumulation of particles smaller than the Kolmogorov scale depending on their inertia and the local flow characteristics has been achieved. Similar maturity for the cases where the feedback of particles in the flow or finite size effects are relevant is still far from being accomplished.\par
The previously mentioned challenges and limitations of experimental and theoretical approaches makes the use of advanced numerical tools a necessity for obtaining detailed information; despite the well-known limitations in terms of Reynolds numbers that can be reached in simulations \citep{Prosperetti-JFM-2015}. Lately, several groups have been successfully using numerical algorithms for interface-resolved direct numerical simulations (DNS) of different turbulent flows laden with finite size particles: examples are suspensions in isotropic turbulence \citep{TenCate-et-al-JFM-2004,Lucci-et-al-JFM-2010}, vertical channel flow \citep{Uhlmann-PoF-2008}, sedimentation \citep{Chouippe-and-Uhlmann-PoF-2015,Fornari-et-al-JFM-2016}, bed load transport \citep{Kidanemariam-and-Uhlmann-JFM-2014,Vowinckel-et-al-AWR-2014}, channel transport of mono-dispersed particles \citep{Wang-et-al-JFE-2016,Yu-et-al-JoT-2016,Wang-et-al-PRF-2017}, and recently of poly-disperse \pc{\citep{Lashgari-et-al-JFM-2016,Fornari-Poly-New}} and non-spherical particles \pc{\citep{Ardekani-et-al-JFM-2016,Eshghinejadfard-AIP-2017}}. Likewise, the present work uses such simulations to study turbulent channel transport of neutrally-buoyant finite size spheres.\par
Suspensions of neutrally-buoyant particles close to the onset of turbulence have also been explored both experimentally \citep{Matas-et-al-PRL-2003} and numerically \citep{Loisel-et-al-PoF-2013}. A common feature of the results from these studies is the fact that solely an increase in effective viscosity of the suspension does not explain the observed phenomenology, in particular for large particles.\par 
In particle suspension flows a new mechanism for momentum transport emerges in the form of a \emph{particle stress} \citep{Batchelor-JFM-1970}. It therefore makes sense to follow the idea of \cite{Lashgari-et-al-PRL-2014}, who distinguished three different flow regimes, depending on the relative importance of viscous, Reynolds and particle stresses to the total stress of the suspension: a laminar (low Reynolds number and low volume fraction), turbulent (high Reynolds number and moderate to low volume fractions) and inertial shear-thickening (high volume fractions), the latest regime being characterized by a significant increase in wall shear that is not accompanied by an increase in the magnitude of the turbulent stresses.\par
Recently, \citet{Picano-et-al-JFM-2015} presented detailed direct numerical simulations of turbulent channel flow laden with neutrally-buoyant finite size spheres. They showed that, for fixed Reynolds number and particle size, particle stresses at a volume fraction of about $20\%$ are responsible for a non-monotonic behavior of the near-wall peak in Reynolds shear stresses as a function of the volume fraction. The associated turbulence attenuation is higher than what predicted only by accounting for an effective viscosity. The decrease in the magnitude of the Reynolds stresses is accompanied by a more dominant increase in particle-induced stresses, which ultimately results in an overall drag increase.\par
This observation is consistent with flow regime map of \cite{Lashgari-et-al-PRL-2014} and supported by the recent work of \citet{Costa-et-al-PRL-2016}. The authors built up upon the work of \cite{Picano-et-al-JFM-2015} by extending the data set to higher Reynolds number and smaller particle sizes. 
They showed that a layer of near-wall particles causes the suspension to deviate from the continuum limit, where its dynamics is well represented by an effective suspension viscosity. Away from this layer, the suspension mean flow is shown to be well described by an effective suspension viscosity. Based on this idea \cite{Costa-et-al-PRL-2016} successfully derived the scaling laws for the mean flow in the overlap region, and accurate correlations capable of predicting the overall drag that the suspension experiences as a function of the three governing parameters: $\Rey_b$, $\Phi$, and $D_p/h$, respectively the bulk Reynolds number, solid volume fraction and particle size ratio.\par
In this work we investigate how finite size effects change the flow and particle dynamics near the wall, and up to which point it is actually important to consider them in the bulk of the channel. We related the observed mesoscale behavior to the local micro-scale dynamics. This allowed us to present a clear picture of the sources of finite size effects in a dense turbulent suspension, needed for future modeling efforts.\par
%
%This paper is organized as follows. In section~\ref{sec:methods} we describe our methodology and computational setup, whereas the results are presented and discussed in section~\ref{sec:results}. The main findings are summarized in section~\ref{sec:conclusions} .

\section{Methods and computational setup}\label{sec:methods}
%\paragraph{Methodology}
%
We performed interface-resolved DNS of a particle suspension flowing through a plane channel in the turbulent regime. The numerical algorithm solves the continuity and Navier-Stokes equations for an incompressible Newtonian fluid with density $\rho$ and kinematic viscosity $\nu$,
\begin{align}
\boldsymbol{\nabla}\cdot\mathbf{u}       &= 0 \mathrm{,} \label{eqn:cont} \\
\frac{\mathrm{D}\mathbf{u}}{\mathrm{D}t} &= -\frac{\boldsymbol{\nabla}{\left(p+p_e\right)}}{\rho} + \nu\boldsymbol{\nabla}^2\mathbf{u}\mathrm{,} \label{eqn:mom}
\end{align}
where $\mathbf{u}$ is the fluid velocity vector, $p+p_e$ the fluid pressure with respect to an arbitrary constant reference value and $\boldsymbol{\nabla} p_e$ corresponds to a constant pressure gradient that may serve as driving force for the flow.\par
This set of equations is solved together with the Newton-Euler equations governing the motion of a spherical particle with mass $m_p$, and moment of inertia $I_p$, 
\begin{align}
m_p \frac{\mathrm{d} \mathbf{U}}{\mathrm{d} t} &= \oint_{\partial V} \! \boldsymbol{\sigma} \cdot \mathbf{n} \, \mathrm{d} A + \mathbf{F}_c \mathrm{,} \label{eqn:nwtn} \\
I_p \frac{\mathrm{d} \boldsymbol{\mathbf{\Omega}}}{\mathrm{d} t} &=\oint_{\partial V} \! \mathbf{r} \times (\boldsymbol{\sigma} \cdot \mathbf{n}) \, \mathrm{d} A + \mathbf{T}_c \mathrm{,} \label{eqn:eulr}
\end{align}
where $\mathbf{U}$ and $\boldsymbol{\Omega}$ denote respectively the particle linear and angular velocity vectors, $\boldsymbol{\sigma} \equiv -(p+p_e) \mathbf{I} + \rho\nu\left(\nabla {\mathbf u} + \nabla {\mathbf u}^T \right)$, $\mathbf{r}$ a position vector with respect to the particle centroid, $\mathbf{n}$ the outward-pointing unit vector normal to the particle surface, $A$ the surface area of the particle, and $\mathbf{F}_c$ and $\mathbf{T}_c$ correspond to external forces and torques associated with short-range inter-particle or particle-wall interactions (such as solid-solid contact).\par
Equations~\eqref{eqn:cont}-\eqref{eqn:mom} and \eqref{eqn:nwtn}-\eqref{eqn:eulr} are coupled through the imposition of no-slip and no-penetration boundary conditions at the particle surface:
\begin{equation}
\mathbf{U} + \boldsymbol{\Omega}\times\mathbf{r} = \mathbf{u} \,\,\,\, \forall \,\,\,\, \mathbf{x} \in \partial V\mathrm{.} \label{eqn:bcs}
\end{equation}
\par A straightforward way of satisfying equation~\eqref{eqn:bcs} is by discretizing the equations in a grid that  conforms to the particles' surfaces -- \pc{a so-called body-fitted method}. Despite the recent progress towards efficient implementation of such methods \cite[see][]{Vreman-JFM-2016}, such implementation remains of prohibitive computational demand for the dense cases considered here. Instead, we use the direct forcing immersed boundary method (IBM) developed in \cite{Breugem-JCP-2012}. The idea of an IBM is that, instead of directly applying the boundary conditions expressed in equation~\eqref{eqn:bcs} by conforming the computational grid to the \pc{surface} of all the particles, one applies a smooth force distribution in strategic regions of the domain such that the boundary condition is satisfied with sufficient accuracy. This way one can benefit from efficient solvers for the Navier-Stokes equations on a regular Cartesian grid.\par 
The fluid flow is solved with a finite-volume pressure correction scheme, coupled with an IBM. Our IBM requires a quasi-2D Lagrangian grid that discretizes the particle surface. The fluid prediction velocity is \emph{interpolated} from the Eulerian to a Lagrangian grid. There the force required for satisfying no-slip and no-penetration is computed. Subsequently, the force is \emph{spread} back to the Eulerian grid. A regularized Dirac delta function with support of 3 grid cells is used to perform interpolation/spreading operations \citep{Roma-et-al-JCP-1999,Uhlmann-JCP-2005}. Regularization of the particle-fluid interface results in first-order spacial accuracy. Slight inward retraction of the Lagrangian grid circumvents this issue, allowing for second-order accuracy in space. Errors due to spreading operations from different neighboring Lagrangian grid points to the same Eulerian grid points are reduced via a multi-direct forcing scheme. We refer to \cite{Breugem-JCP-2012} for more details on this IBM and Navier-Stokes solver.\par
Short-range hydrodynamic particle-particle and -wall interactions are reproduced by the IBM when the gap sizes are sufficiently well resolved. However, since the method uses a fixed grid, a lubrication model is inevitably needed once the inter-particle gap width becomes smaller than the grid spacing. 
%Our lubrication model is based on asymptotic expansions of analytical solutions for canonical lubrication interactions between spheres in the Stokes regime. Roughness effects are incorporated by making the lubrication correction independent of the gap width for gap widths smaller than $\sim 1\%$ of the particle radius. This correction is applied until the particles reach solid-solid contact. Then, the corresponding contact force is modeled by a soft-sphere collision model. 
The closures for short-range particle-particle and particle-wall interactions used in the present study \pc{have been validated in} \cite{Costa-et-al-PRE-2015}, \pc{and are summarized below}.\par
\pc{Normal lubrication interactions unresolved by the IBM, and roughness effects are accounted for by a two-parameter model. When a sphere approaches a wall or another sphere, reaching a normalized gap distance $\varepsilon_{\Delta x}$ (with $\varepsilon\equiv \delta_{ij,n}/R_p$, with $\delta_{ij,n}$ being the gap distance between particles $i$ and $j$), a lubrication correction force is computed. The parameter $\varepsilon_{\Delta x}$ corresponds to the non-dimensional threshold distance below which the IBM cannot resolve canonical lubrication interactions in the Stokes regime. The closure uses a correction force $\Delta F_{lub}=-6\pi\mu R_p u_{ij,n}(\lambda(\varepsilon)-\lambda({\varepsilon_{\Delta x}}))$, added to the integration of the particle momentum equation~\eqref{eqn:nwtn}, where $\lambda$ is the Stokes amplification factor. To account for rougnhess effects, this correction is made independent of the gap-width for $0<\varepsilon<\varepsilon_\sigma$: $\Delta F_{lub} \propto u_{ij,n}(\lambda(\varepsilon_\sigma)-\lambda{\varepsilon_{\Delta x}})$. When the two surfaces start to overlap, the lubrication correction is set to zero and a soft-sphere collision model takes over. These corrections are based on asymptotic expansions of analytical solutions for short-range particle-particle and particle-wall interactions in the Stokes regime, see e.g.\ \cite{Dance-and-Maxey-JCP-2003} for a review.}\par
\pc{The soft-sphere model computes the normal collision force from the following linear spring-dashpot model:}
\begin{equation}
  \mathbf{F}_{ij,n} = -k_n\boldsymbol{\delta}_{ij,n}-\eta_n\mathbf{u}_{ij,n}
\end{equation}
\pc{where $\boldsymbol{\delta}_{ij}$ and $\mathbf{u}_{ij,n}$ are the gap-width and relative particle velocity projected in the line-of-centers $\mathbf{n}_{ij}=(\mathbf{x}_j-\mathbf{x}_i)/(||\mathbf{x}_j-\mathbf{x}_i||)$; $\mathbf{x}_{i/j}$ corresponds to the centreline position of particles $i/j$. The spring and dashpot coefficients are given by:}
\pc{\begin{equation}
k_n = \frac{m_e\left(\pi^2+\ln^2 e_{n,d}\right)}{(N\Delta t)^2}\mathrm{,} \; \eta_n = -\frac{2m_e\ln e_{n,d}}{(N\Delta t)}\mathrm{,}
\end{equation}
where $e_{n,d}$ is the dry coefficient of restitution, and $m_{e} = \left(m_i^{-1}+m_j^{-1}\right)^{-1}$ the reduced mass of the particles. $N\Delta t$ is the collision time, set as a multiple $N$ of the time step of the Navier-Stokes solver, $\Delta t$. This allows the flow solution to gradually adapt to the sudden changes in particle velocity \citep{Costa-et-al-PRE-2015,Biegert-et-al-JCP-2017}. The model for the tangential component of the collision force in \citet{Costa-et-al-PRE-2015} is not described here, as in the present study we consider the particles to be frictionless; as discussed later, frictional interactions can be neglected for the flow governing parameters at stake. The values of $\varepsilon_{\Delta x}$, $\varepsilon_\sigma$ and $e_{n,d}$ used are given in table~\ref{tbl:short_range_params}.}
\pc{
\begin{table}
 \centering
 \begin{tabular}{l c c c c}
 interaction       & $e_{n,d}$ & $\varepsilon_{\Delta x}$ & $\varepsilon_\sigma$ & $N$ \\
 particle-wall     & $0.97$      & $0.075$                  & $0.001$              & $10$ \\
 particle-particle & $0.97$      & $0.025$                  & $0.001$              & $10$ \\
 \end{tabular}
 \caption{Parameters for lubrication and contact models for short-range particle-particle and particle-wall interactions.}\label{tbl:short_range_params}
\end{table}
}
\pc{The parameters in table~\ref{tbl:short_range_params} are taken from \cite{Costa-et-al-PRE-2015}, where a thorough validation is performed against several canonical cases, like the trajectory of a sphere colliding onto a planar surface in a viscous liquid, and wet coefficients of restitution for head-on particle-wall and particle-particle collisions.}\par
%\todo{Define here $e_n$, $\varepsilon$, $N$ and justify why frictionless particles are okay!.}}
%
%{\color{blue}The numerical algorithm uses a 2D domain decomposition, where both the fluid field and particles are distributed into \textit{pencils}. For simplicity of implementation, the sizes of these computational subdomains are restricted by the particle size. Whenever this restricts the number of cores, a hybrid distributed/shared memory parallelization is used.}\par
%
The flow dynamics is governed by three parameters: the bulk Reynolds number $\Rey_b\equiv U_b(2h)/\nu$, the particle size ratio $D_p/h$, and the bulk volume fraction of solid particles $\Phi = N_pV_p/V_t$ (note that the particles are neutrally-buoyant), where $U_b$ is the flow bulk velocity (forced to be constant in the numerical algorithm), $h$ the half channel height, $D_p$ the particle diameter, $N_p$ the total number of particles, and $V_p$ and $V_t$ the volumes of a particle and of the computational domain. In the present work the bulk volume fraction of solid particles is fixed to $\Phi=20\%$, and the Reynolds number to $\Rey_b=12\,000$, in order to ensure sufficient inner-to-outer scale separation \cite{Costa-et-al-PRL-2016}. Two different particle sizes are considered as reported in table~\ref{tbl:comp_params}, together with other relevant physical and computational parameters.\par
\pc{We explore the dataset of dense suspension flow reported in \citep{Costa-et-al-PRL-2016}}. The simulations were performed in a domain with dimensions $L_x/h\times L_y/h\times L_z/h=6\times 2 \times 3 $, using a grid spacing dictated by the number of grid cells required to resolve the flow conforming the spheres: $\Delta x/D_p = 1/16$\pc{, which results in a distribution of $746$ Lagrangian grid points on the surface of each particle}. The interface-resolved simulations were complemented with two single-phase reference cases: the unladen case at the same bulk Reynolds number, denoted SPR (single-phase reference), and the continuum limit where the flow dynamics can be reproduced by a single-phase fluid with the effective viscosity of the suspension at $\Phi=20\%$, denoted as CLR (continuum limit reference). This simulation corresponds thus to a single-phase flow with  bulk Reynolds number $\Rey_b\nu/\nu^e\approx 6\,400$ and effective viscosity $\nu^e = \nu (1+(5/4)\Phi/(1-\Phi/\Phi_{max}))^2$ obtained from Eilers fit \cite{Stickel-and-Powell-ARFM-2005}, with $\Phi_{max}=0.6$.\par
\pc{Some remarks on the choice of the collision model parameters should be made at this point. First, the choice of $e_{n,d}=0.97$ is motivated by values measured experimentally for dry collisions of rigid spheres, \citep{Foerster-et-al-PoF-1994,Joseph-et-al-JFM-2001}. However, the results should not be sensitive to small changes in $e_{n,d}$. The reason is that, as we will see, the particle inertia is sufficiently small that the resulting impact Stokes number $\mathrm{St}\equiv (1/9)\rho_p u_{ij,n} D_p/\mu < 10$. In this scenario, hydrodynamic viscous effects dominate the collision dynamics and the particles typically do not rebound, despite the value $e_{n,d}\approx1$ \citep{Legendre-et-al-CES-2006,Yang-and-Hunt-PoF-2006}. Second, the effect of particle-particle and particle-wall solid friction in the suspension dynamics is small. For particle-particle interactions, the mean separation distance ($\sim \Phi^{-1/3} D_p$) is sufficiently large that sustained frictional contacts are unlikely. In regard to particle-wall interactions, there is no mechanism causing a large net force for neutrally-buoyant particles, pushing them towards the wall. Since the frictional force is proportional to this normal force, solid-solid friction at the wall is negligible. In conclusion, the main role of the collision model in this flow is ensuring that particles occupy their own volume; i.e.\ once particles are in the imminence of contact, they cannot overlap. Then the solid-fluid coupling dictate the flow dynamics. Finally, we do not implement lubrication corrections in the tangential direction. These choices are supported by the success of this approach in reproducing the effective viscosity of a suspension up to $\Phi=30\%$ in \cite{Picano-et-al-PRL-2013}, where friction and tangential lubrication corrections were neglected.}\par% Still, we confirmed this in simulations of laminar shear flow with neutrally-buoyant particles at $\Phi=20\%$ (same configuration as in \citet{Picano-et-al-PRL-2013}) that including friction (Coulomb coefficient of sliding friction $\mu_s=0.2$) resulted in an increase in effective viscosity of $0.X\%$.}\par

%Finally, we do not implement lubrication corrections in the tangential direction. In contrast with the normal lubrication force, which scales as $1/\varepsilon$ ($\varepsilon\equiv$ gap-width normalized by the particle radius), the tangential force and torque scale with $\ln(\varepsilon)$, and thus diverges much slower. Hence, the fraction of the total force unresolved by the IBM at grid scales $\sim \Delta x$ is much smaller. 

%
\pc{Unless otherwise stated, spatial averages of an observable $o$ pertaining to a certain phase are obtained from the following intrinsic volume average:
\begin{equation}
\left<o\right>^{f/s}(y) = \frac{\sum_{ik}o_{ijk}(x,y,z)\phi_{ijk}^{f/s}(x,y,z)}{\sum_{ik}\phi_{ijk}^{f/s}(x,y,z)}
\end{equation}
where $\phi(x,y,z)$ is a phase-indicator function ($\phi_{ijk}^{f} + \phi_{ijk}^{s} = 1$), and the superscripts $f$ and $s$ denote fluid and solid, respectively. For simplicity of notation we will drop the brackets $\left<\right>$. Profiles pertaining to the combined phase are obtained similarly,
\begin{equation}
\left<o\right>^{c}(y) = \frac{\sum_{ik}\left(o_{ijk}(x,y,z)\phi_{ijk}^{f}(x,y,z)+o_{ijk}(x,y,z)\phi_{ijk}^{s}(x,y,z)\right)}{\sum_{ik}\left(\phi_{ijk}^{f}(x,y,z)+\phi_{ijk}^{s}(x,y,z)\right)}\mathrm{.}
\end{equation}
When computing the profiles of particle velocity, the Eulerian grid points in the solid domain are populated by the velocity corresponding to rigid-body motion: $\mathbf{U}_p(x,y,z) = \mathbf{U} + \boldsymbol{\Omega}\times(\mathbf{x}(x,y,z)-\mathbf{x}_c)$, with $\mathbf{x_c}$ the particle centroid position.}\par
Due to the large size of the simulations, all spatial averages were computed during run time and averaged in time a posteriori, with samples obtained from a time interval of about $1500h/U_b$, spaced by $7h/U_b$. A total of about 7 million core hours in the supercomputer CURIE (thin nodes; B510 bullx) at CEA, France, and Beskow (Cray XC40) at KTH, Sweden were used.
\begin{table}
 \centering
 \begin{tabular}{l r r r r r r}
 Case   & $\Rey_b$ & $h/D_p$ & $D_p/\delta_v^{sph}$ & $\Phi\, (\%)$ & $N_x \times N_y  \times N_z $ & $N_p     $ \\
 D10    & $12\,000$ & $36$ & $9.7 $ & $20$ & $3456\times 1152 \times 1728$ & $640\,000$ \\
 D20    & $12\,000$ & $18$ & $19.4$ & $20$ & $1728\times ~576 \times ~864$ & $~80\,000$ \\
 SPR    & $12\,000$ & --   & --     & --   & $1728\times ~576 \times ~864$ & $~~~~~~~0$ \\
 CLR    & $~6\,400$ & --   & --     & --   & $1152\times ~384 \times ~576$ & $~~~~~~~0$
 \end{tabular}
 \caption{Physical and computational parameters of the DNS data sets. $\delta_v^{sph}$ ($\gtrsim \delta_v$) denotes the viscous wall unit for the corresponding single-phase flow at the same $\Rey_b$, estimated from $\Rey_\tau^{sph}=0.09\Rey_b^{0.88}$ \citep{Pope-2001}. $N_p$ denotes the number of particles, $N_{x/y/z}$ denotes the number of grid points in the $x/y/z$ directions.}\label{tbl:comp_params}
\end{table}

\section{Results}\label{sec:results}
\subsection*{Near-wall dynamics -- the main cause of finite size effects}
We will here refer to \textit{finite-size effects} as the causes for the meso or macroscale-averaged flow statistics to be different from those obtained from the continuum limit approximation in which the suspension is modeled as a Newtonian fluid with an effective viscosity due to the presence of the particles.\par
\cite{Costa-et-al-PRL-2016} showed that a layer of particles, flowing near the wall with significant (apparent) slip velocity, is responsible for an additional increase in drag which cannot be modeled by an effective suspension viscosity. By accounting for this effect, the authors were able to scale the profiles of mean velocity and velocity defect ($U_c-u$, where $U_c$ is the mean centerline velocity) for a wide range of combinations of $\Rey_b$, $\Phi$ and $D_p/h$. The theory in \cite{Costa-et-al-PRL-2016} assumes that the domain can be split into two regions, bounded at a wall-normal distance $y=\delta_{pw}$: a region away from the wall \pc{with uniform particle concentration}, denoted homogeneous suspension region (HSR) ($y>\delta_{pw}$) where the mean flow of the suspension is well represented by the continuum limit of a Newtonian fluid with effective viscosity $\nu^e$, and a region close to the wall denoted particle-wall layer (PWL) ($y<\delta_{pw}$) where the difference between the mean flow of the two phases, at the mesoscale level, makes this assumption invalid. By exploiting the stress budget, the authors could derive scaling laws for the mean velocity and velocity defect in the HSR, provided that, like in a single-phase turbulent channel, inner-to-outer scale separation is satisfied. In other words, the friction Reynolds number based on the scaling parameters of the homogeneous suspension region needs to be sufficiently high. These are a channel height corrected for a virtual-wall origin, $h-\delta_{pw}$, the friction velocity taken from the profile of the total stresses at $y=\delta_{pw}$, $u_\tau^*=\sqrt{u_\tau^2(1-\delta_{pw}/h)}$ and the effective suspension viscosity $\nu^e$, yielding $\Rey_\tau^{e*} = u_\tau^*(h-\delta_{pw})/\nu^e$. The relations derived in \cite{Costa-et-al-PRL-2016} read,
\begin{align}
  \frac{u}{u_\tau^*} = &\frac{1}{\kappa} \ln\left(\frac{y-\delta_{pw}}{\delta_v^{e*}}\right) + B \mathrm{,} \label{eqn:scaling_um} \\
  \frac{U_c - u}{u_\tau^*} = - &\frac{1}{\kappa} \ln\left(\frac{y-\delta_{pw}}{h-\delta_{pw}}\right) + B_d \mathrm{,} \label{eqn:scaling_ud}
\end{align}
with $\delta_v^{e*} = \nu^e/u_\tau^*$;  the coefficients $\kappa$, $B$ and $B_d$ retain the values of single-phase flows, whereas the effective viscosity is obtained from Eilers fit $\nu^e/\nu = (1+(5/4)\Phi/(1-\Phi/\Phi_{max}))^2$ \citep{Stickel-and-Powell-ARFM-2005}. 
The virtual wall origin of the HSR is given by $y=\delta_{pw}=C(\Phi/\Phi_{max})^{1/3}D_p$. These scaling laws can be further used to derive a master equation that accurately predicts the overall drag of the suspension, here expressed in terms of a friction Reynolds number:
\begin{table}
 \centering
 \begin{tabular}{l r r r}
 Case          & $\Rey_\tau$ & $\Rey_\tau^e$ & $\Rey_\tau^{e*}$ \\
 SPR (unladen) & $350             $ & $350               $ & $350                  $ \\
 D10           & $395             $ & $209               $ & $203                  $ \\
 D20           & $406             $ & $215               $ & $203                  $ \\
 CLR           & $201             $ & $201               $ & $201                  $ \\
 \end{tabular}
 \caption{Friction Reynolds numbers for the different cases studied. $\Rey_\tau=u_\tau h/\nu$ is the typical friction Reynolds number defined from the wall friction velocity and fluid viscosity, $\Rey_\tau^e=\Rey_\tau\nu/\nu^e$ is instead defined from the suspension effective viscosity and $\Rey_\tau^{e*}=u_\tau(1-\delta_{pw}/h)^{1/2}(h-\delta_{pw})/\nu^e=\Rey_\tau^e(1-\delta_{pw}/h)^{3/2}$ is defined from a wall friction velocity and channel height corrected for finite-size effects and the effective suspension viscosity.}\label{tbl:frictionre}
\end{table}
\begin{equation} 
\Rey_\tau =  \frac{\Rey_b}{2\xi_{pw}^{1/2}}\left(\frac{1}{\kappa}\left[\ln\left(\Rey_\tau\chi^e\xi_{pw}^{3/2}\right)-1\right]+B+B_d\right)^{-1}\mathrm{,}\label{eqn:scaling_ret}
\end{equation}
or, in an explicit form based on a similar correlation for single-phase flow from \citet{Pope-2001},
\begin{equation}
  \Rey_\tau = \frac{0.09\left(\Rey_b\chi^e\xi_{pw}\right)^{0.88}}{\xi_{pw}^{3/2}\chi^e}\mathrm{,} \label{eqn:retau_empirical}
\end{equation}
where $\xi_{pw} = (1-\delta_{pw}/h)$ and $\chi^e=\nu/\nu^e$.\par
Table~\ref{tbl:frictionre} presents the mean wall shear, expressed in terms of the friction Reynolds number $\Rey_\tau = u_\tau h/\nu$ from the different simulations considered here. As expected from eqs.~\eqref{eqn:scaling_ret} and \eqref{eqn:retau_empirical}, the addition of finite size neutrally-buoyant particles results in an increase in drag with respect to the value of 350 of the single phase flow, which (when fixing the other governing parameters) increases with increasing particle size. 
We first note that the smaller the particles, the more the \emph{suspension} friction Reynolds number $\Rey_\tau^e = \Rey_\tau\nu/\nu^e$ approaches the value obtained in the continuum limit, CLR; in other words using an effective viscosity provides a better prediction of the total drag.
This is a consequence of the reduced finite-size effects that typically occur near the wall \citep{Costa-et-al-PRL-2016}. \par
Finally, the third column in the table shows the suspension Reynolds number of the HSR, $\Rey_\tau^{e*}=\Rey_\tau^e(1-\delta_{pw}/h)^{3/2}$, obtained considering both an effective suspension viscosity and a virtual wall origin as explained above (with $\delta_{pw}$ computed with $C=1.5$ and $\Phi_{\max}=0.6$). The friction Reynolds numbers from the  interface-resolved simulations are, in this case, equal for both particle sizes and very close to the value of $\Rey_\tau\nu/\nu^e$ for the CLR case. This strongly supports the proposed correction for finite-size effects.\par 
\begin{figure}
  \centering
  \includegraphics[width=0.46\textwidth]{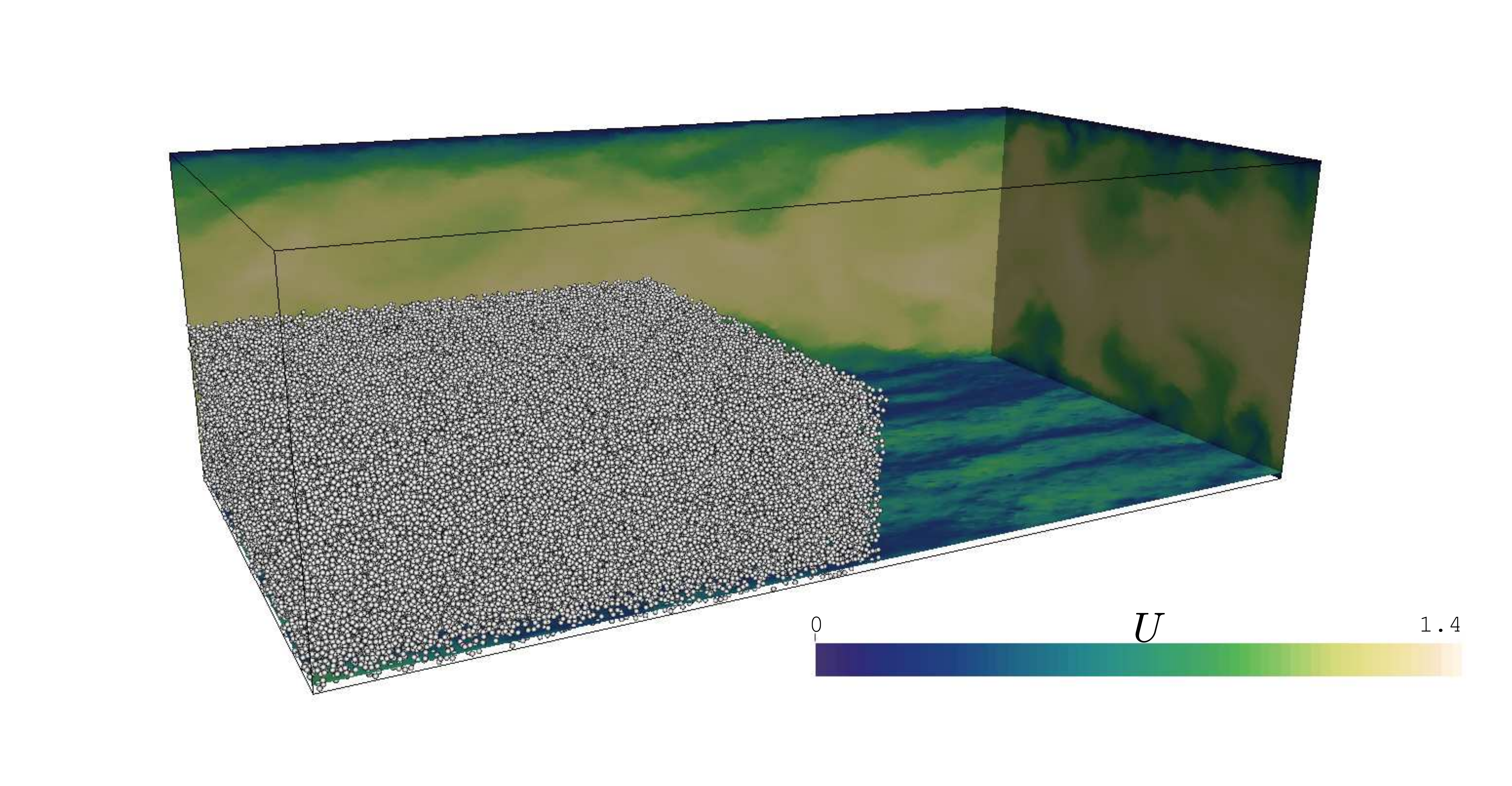}(a)\hfill
  \includegraphics[width=0.46\textwidth]{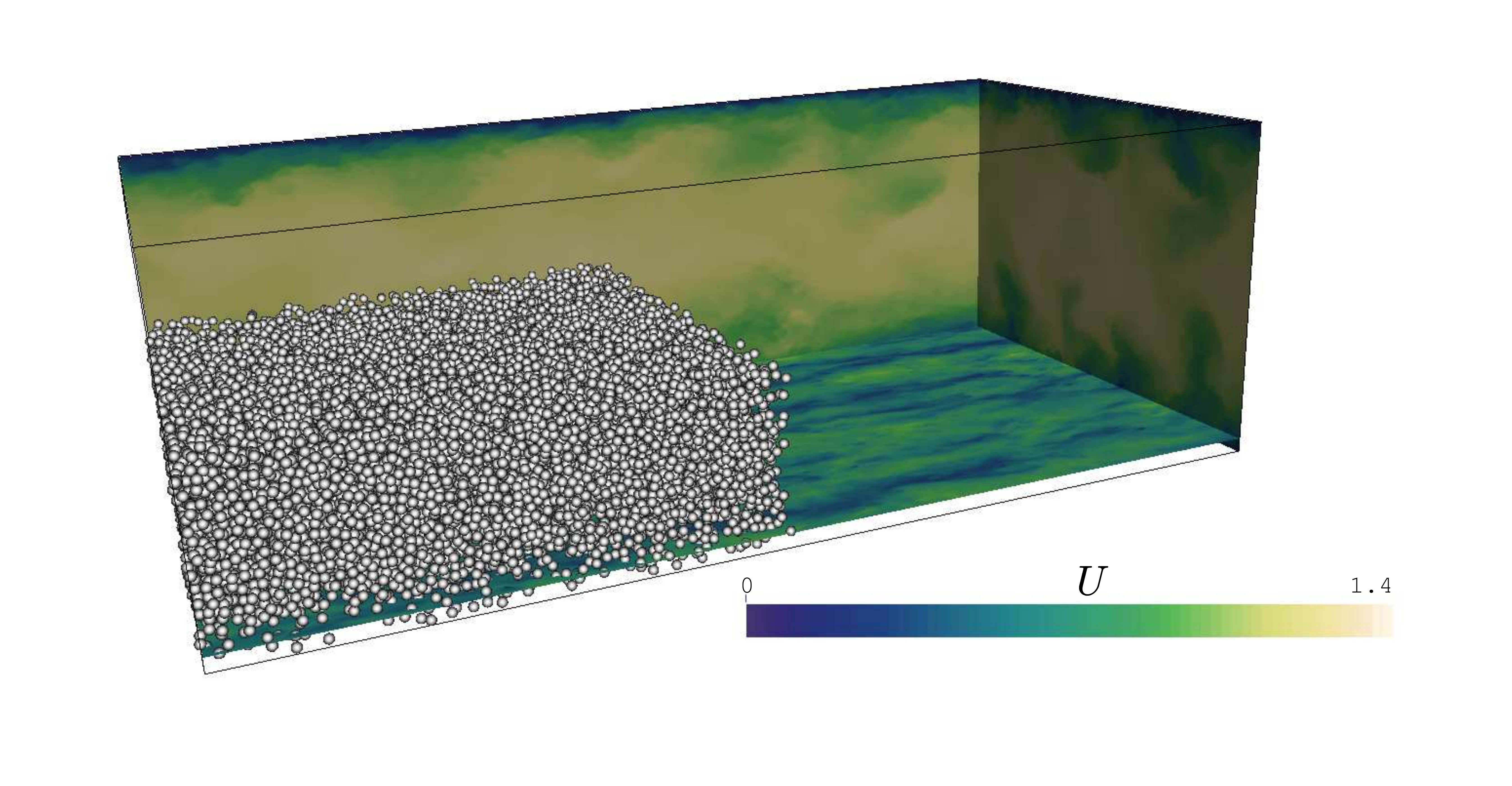}(b)\\
  \includegraphics[width=0.46\textwidth]{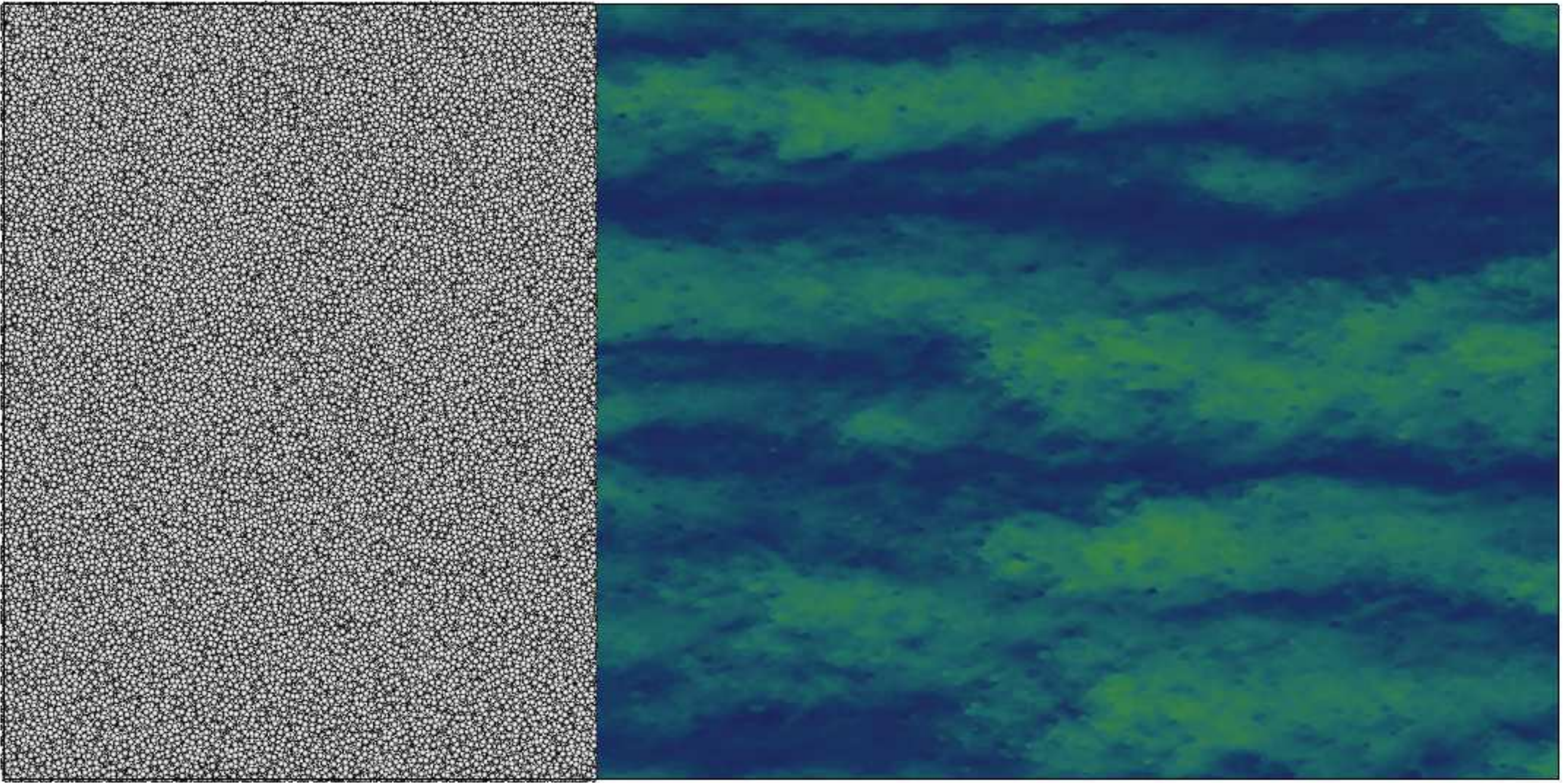}(c)\hfill
  \includegraphics[width=0.46\textwidth]{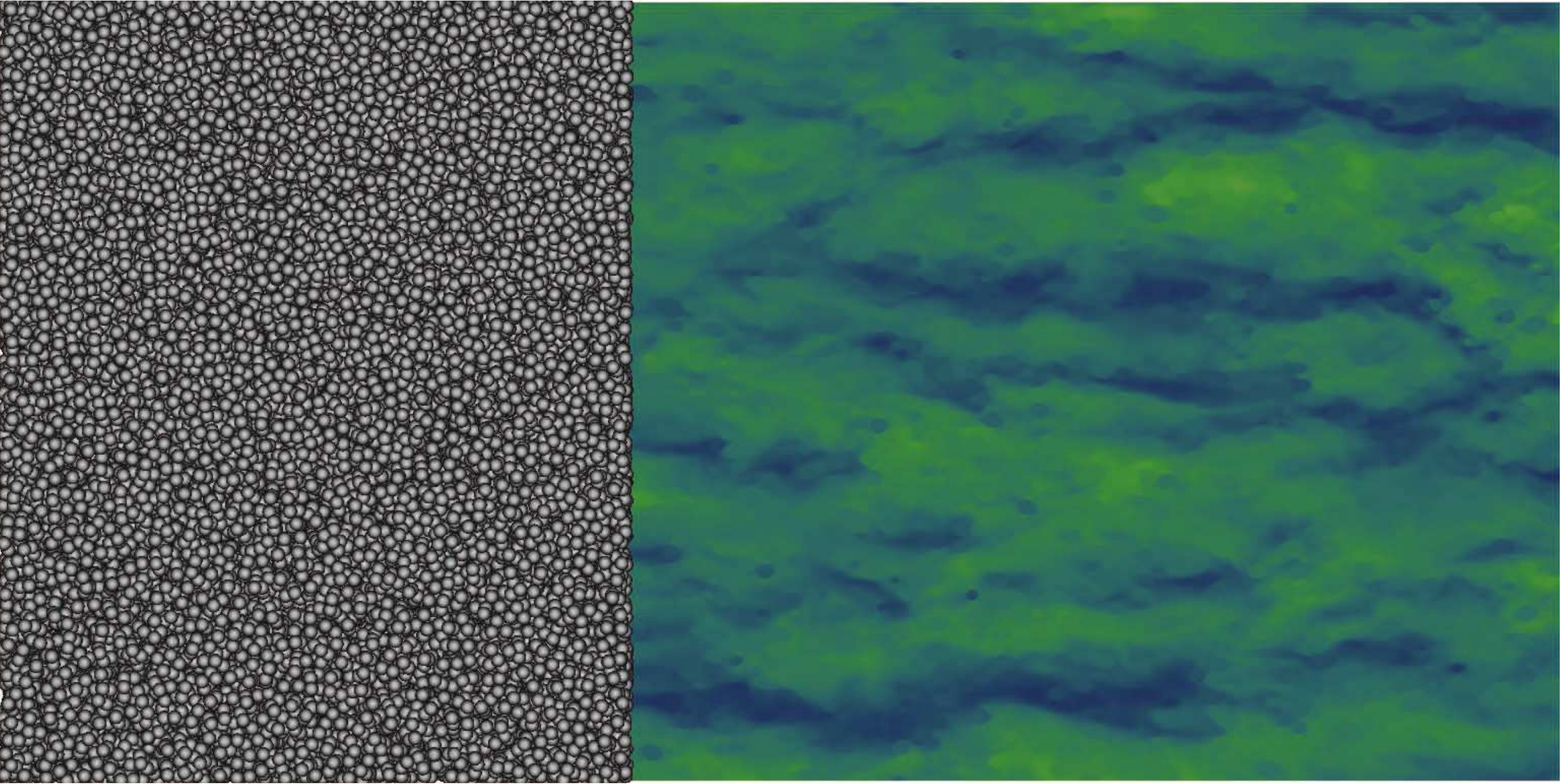}(d)
  \caption{Instantaneous snapshots of the flow for cases D10 (a) and D20 (b). The contours show the of magnitude of streamwise velocity (normalized by $U_b$) in three mutually-perpendicular planes. Particles are illustrated in part of the domain. The bottom plane corresponds to a wall-normal distance of $y=1.5 D_p$, also shown in a 2D perspective in panels (c) (D10) and (d) (D20).}
\label{fig:visus}
\end{figure}
\begin{figure}
  \centering
  \includegraphics[width=0.99\textwidth]{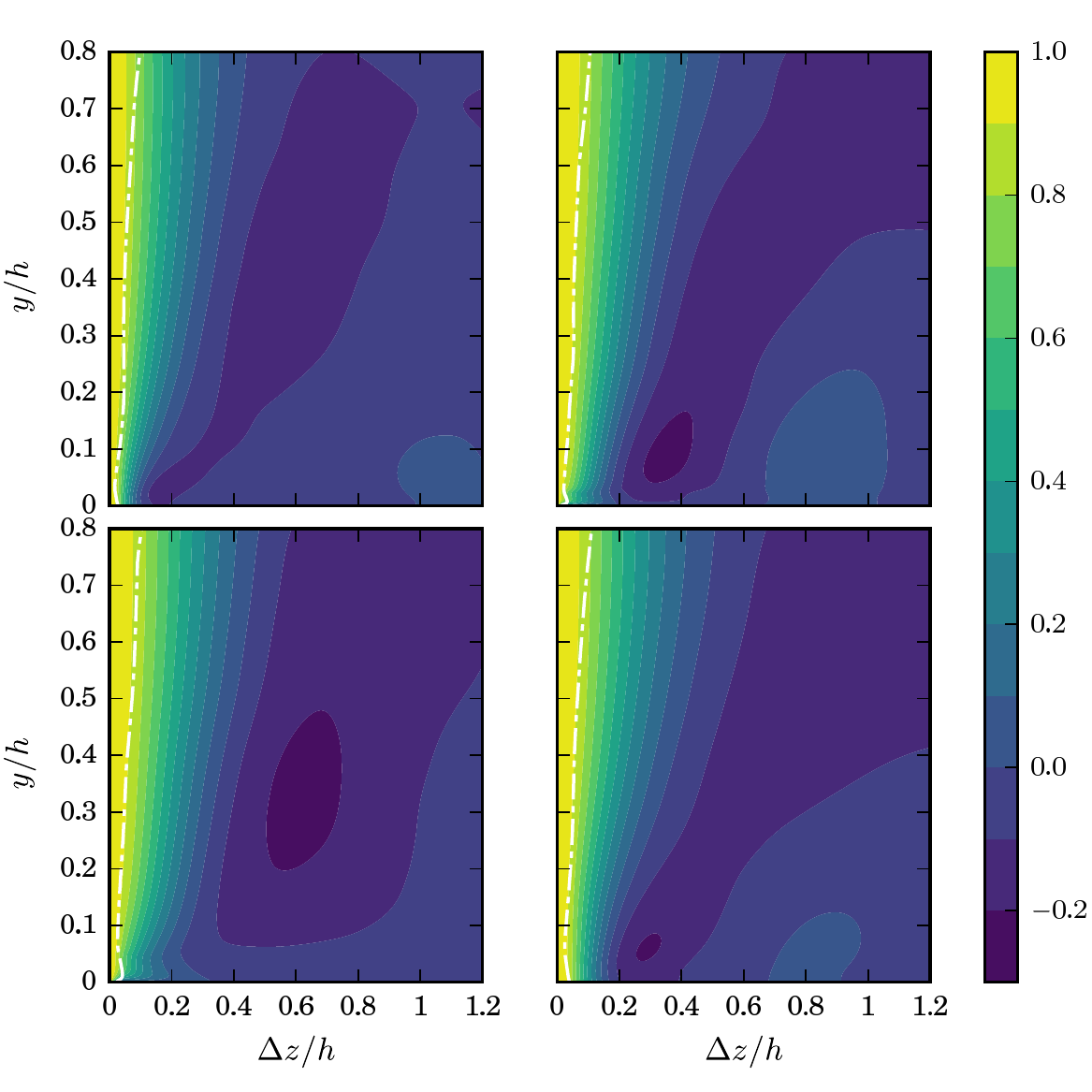}\hfill
  \put(-75 ,50 ){\color{white}(d)}
  \put(-230,50 ){\color{white}(c)}
  \put(-75 ,215){\color{white}(b)}
  \put(-230,215){\color{white}(a)}
  \caption{Autocorrelations of streamwise velocity $R_{uu}^z$ as a function of the wall-normal distance, $y$, for spanwise separation distances $\Delta z$ (both scaled with $h$) for cases SPR (a), D10 (b), D20 (c) and CLR (d). The dashed lines denote the profile of integral scale $L_{xz}(y)$.}
\label{fig:autocorrs}
\end{figure}

Figure~\ref{fig:visus} displays snapshots of the flow for the two laden cases, D10 and D20. The top panels show planar sections colored by the streamwise flow velocity, and particle positions only shown for streamwise locations smaller than $0.4L_x$ and wall-normal distance $y < h$. The bottom plane (also shown in the two planar views, in the bottom panels) corresponds to a wall-normal position $y=1.5 D_p$, corresponding to about $y/\delta_v=15$ and $30$ for cases D10 and D20. The figures show the typical near-wall low-speed streaks found in single-phase wall-bounded turbulence, also present in these suspensions \citep{Picano-et-al-JFM-2015}. Differences in velocity contrast between the two cases can be attributed to the choice of a plane at wall-normal distance that scales with the particle diameter, meaning that, in viscous units, the plane corresponding to case D10 is closer to the wall. However, in both cases the maximum streak amplitude is achieved at relatively larger wall-normal distances than what is expected for single-phase flows, as illustrated in figure~\ref{fig:autocorrs}.
This figure depicts contours of autocorrelations of the streamwise suspension velocity (i.e.\ computed from the fluid flow field with rigid body motion inside the particles) $R_{uu}^z(y,\Delta z) \equiv \left<u'(y,z)u'(y,z+\Delta z)\right>/\left<u'^2(y)\right>$, averaged in time and in the streamwise direction for different spanwise separations.
Panels (b) and (c) show the results for the interface-resolved simulations D10 and D20. Clearly the near-wall minima -- footprint of the low- and high-speed streaks \citep{Kim-et-al-JFM-1987}, and near-wall property that should scale in inner units -- are shifted upwards, and the larger the particles are, the larger this shift is.\par
Figure~\ref{fig:autocorrs} also shows that the smaller the particles are, the closer the suspension dynamics resemble the continuum limit, CLR. \pc{This can be seen when comparing panel (d), corresponding to the continuum limit CLR to panels (b) -- D10 and (c) -- D20. There is a clear qualitative discrepancy between the CLR and the bigger particles (D20), contrasting with a significant quantitative agreement between CLR and the case with small particles (D10).} For case D20, the wall-normal minimum is located at such a large distance to the wall ($\approx 0.3h$) that inner-to-outer scale separation is compromised. Also, the autocorrelations for the interface-resolved cases show a non-monotonic trend at wall-normal distances $y\lesssim D_p$, where the autocorrelation (for fixed $y$) reaches a local maximum close to the wall. This can be even better observed from the profile of the integral length scale $L_{xz}=\int_{-L_z/2}^{+L_z/2} R_{uu}^z\,\mathrm{d}(\Delta z)$, indicated by the dashed lines in the figure. This confirms that the flow dynamics in this region is qualitatively different from what a simple continuum rheological description would predict.

\subsection*{Flow dynamics near the particle-wall layer (PWL)}
\begin{figure}
\centering
\includegraphics[width=0.33\textwidth]{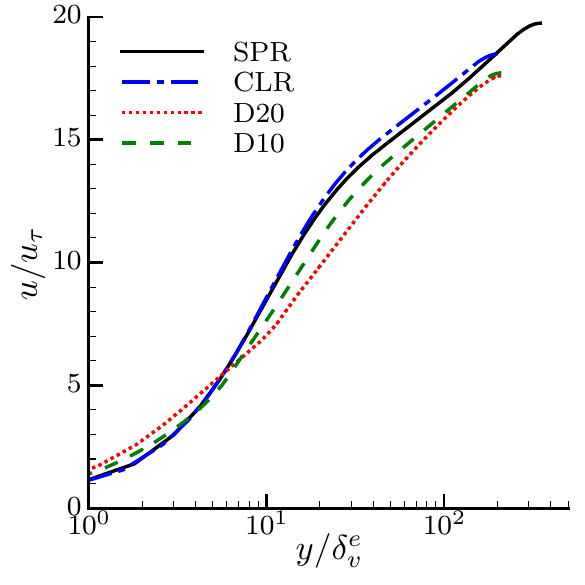}\hfill
\includegraphics[width=0.33\textwidth]{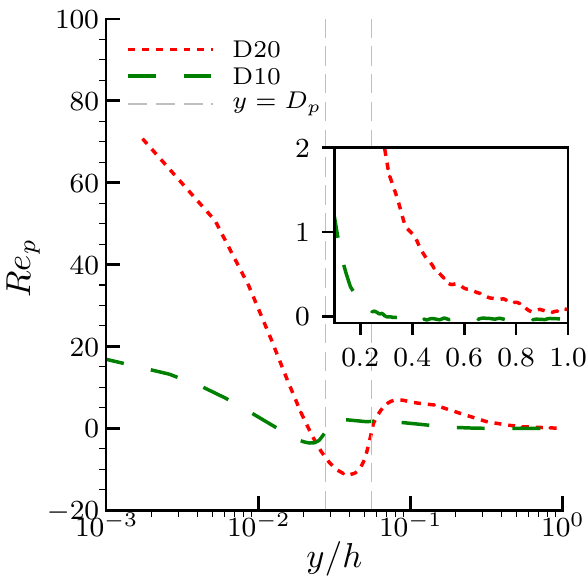}\hfill
\includegraphics[width=0.33\textwidth]{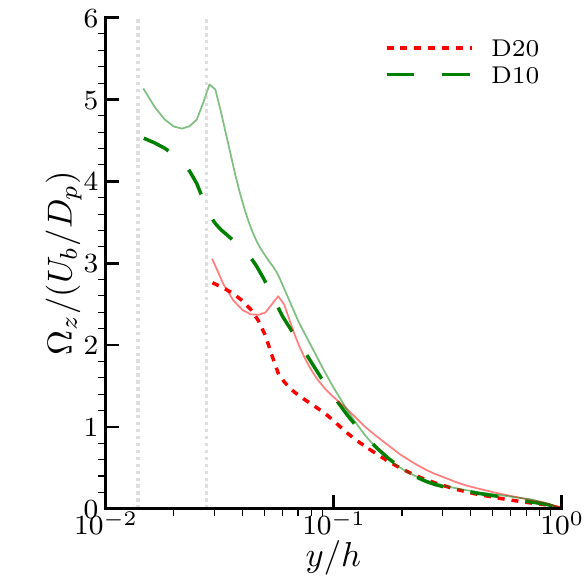}
%  \put(-25 ,40 ){\color{black}(b)}
%  \put(-220,40 ){\color{black}(a)}
  \put(-280,24){\color{black}(a)}
  \put(-150,24){\color{black}(b)}
  \put(-20 ,24){\color{black}(c)}
%\caption{(a)  Mean streamwise (inner-scaled) fluid velocity and (b) (outer-scaled) particle velocity. In panel (b) the fluid velocity is also shown in thin grey lines, while vertical dashed lines indicate the wall-normal distance corresponding to one particle diameter.}\label{fig:mean_vel}
\caption{(a) Mean streamwise (inner-scaled) fluid velocity and (b) profile of particle Reynolds number based on the (apparent) particle-to-fluid slip velocity $Re_p=(u_p-u_f)D_p/\nu$. Vertical dashed lines indicate the wall-normal distance corresponding to one particle diameter. The inset shows the same quantities, but with a linear-linear scale, to highlight the differences in particle Reynolds number in the bulk. \pc{Note that both translation and rotation are accounted for when computing $u_p$. (c) Fluid and particle angular velocity versus the outer-scaled wall-normal distance. The solid lines depict the fluid velocity, and the vertical dotted lines denote the wall-normal distance corresponding to $y=R_p$ for both cases.}}\label{fig:mean_vel}
\end{figure}
Next we investigate in more detail the flow dynamics near the particle-wall layer. 
Panel (a) of figure~\ref{fig:mean_vel} depicts the mean streamwise fluid velocity in inner units. Note that the viscous wall unit used for inner-scaling is defined, for consistency, with the effective suspension viscosity at the same volume fraction, as we are ultimately interested in the deviations of the flow dynamics from the continuum limit.
For wall-normal distances $y \gtrsim 10\delta_v^e$, the profile for the case D10 (smaller particles) clearly shows a  logarithmic scaling, with a  von K\'arm\'an constant $\kappa = 0.36$. On the other hand, case D20 does not show a clear logarithmic region because of the larger extent of the PWL \citep{Costa-et-al-PRL-2016}.\par
Panel(b) of figure~\ref{fig:mean_vel} presents the difference between the fluid- and solid-phase velocity profiles, expressed in terms of a particle Reynolds number $Re_p=(u_p-u_f)D_p/\nu$. Profiles pertaining to the solid phase are obtained by averaging over the rigid body motion of the particles. Two regions can be clearly distinguished in the figure, roughly separated by the line marking the wall-normal distance $y=D_p$. For $y\gtrsim D_p$, the difference between the velocity of the two phases is small, whereas the profiles clearly deviate for wall-normal distances $y\lesssim D_p$, reaching the highest (apparent) particle-to-fluid slip velocity at the wall. This is a signature of particle layering due to the kinematic constraint that the wall imposes on the particles. Particles flowing at the wall acquire most of their linear momentum at wall-normal distances higher than their radius and this is transported uniformly throughout their volume. Conversely, the fluid momentum must vanish at the wall. Notably, Figure~\ref{fig:mean_vel}~(b) shows (see also its inset) that the particle Reynolds number in the bulk is virtually zero for case D10, while it still assumes values of $O(1)$ for case D20. This evidence of finite particle inertia in the bulk can result in inertial shear-thickening effects \cite[see][]{Picano-et-al-PRL-2013}. We will come back to this when the dynamics of the homogeneous suspension region are discussed. \pc{Note that in this figure both the effect of particle translation and rotation is accounted for. To discern the effect of rotation, panel (c) of figure ~\ref{fig:mean_vel} compares the profile of particle and fluid angular velocity for the two cases. A similar trend is observed: larger deviations with respect to the fluid profile is found for larger particles, and the highest discrepancy between fluid and particle statistics occurs close to the wall.}\par
\begin{figure}
\centering
\includegraphics[width=0.99\textwidth]{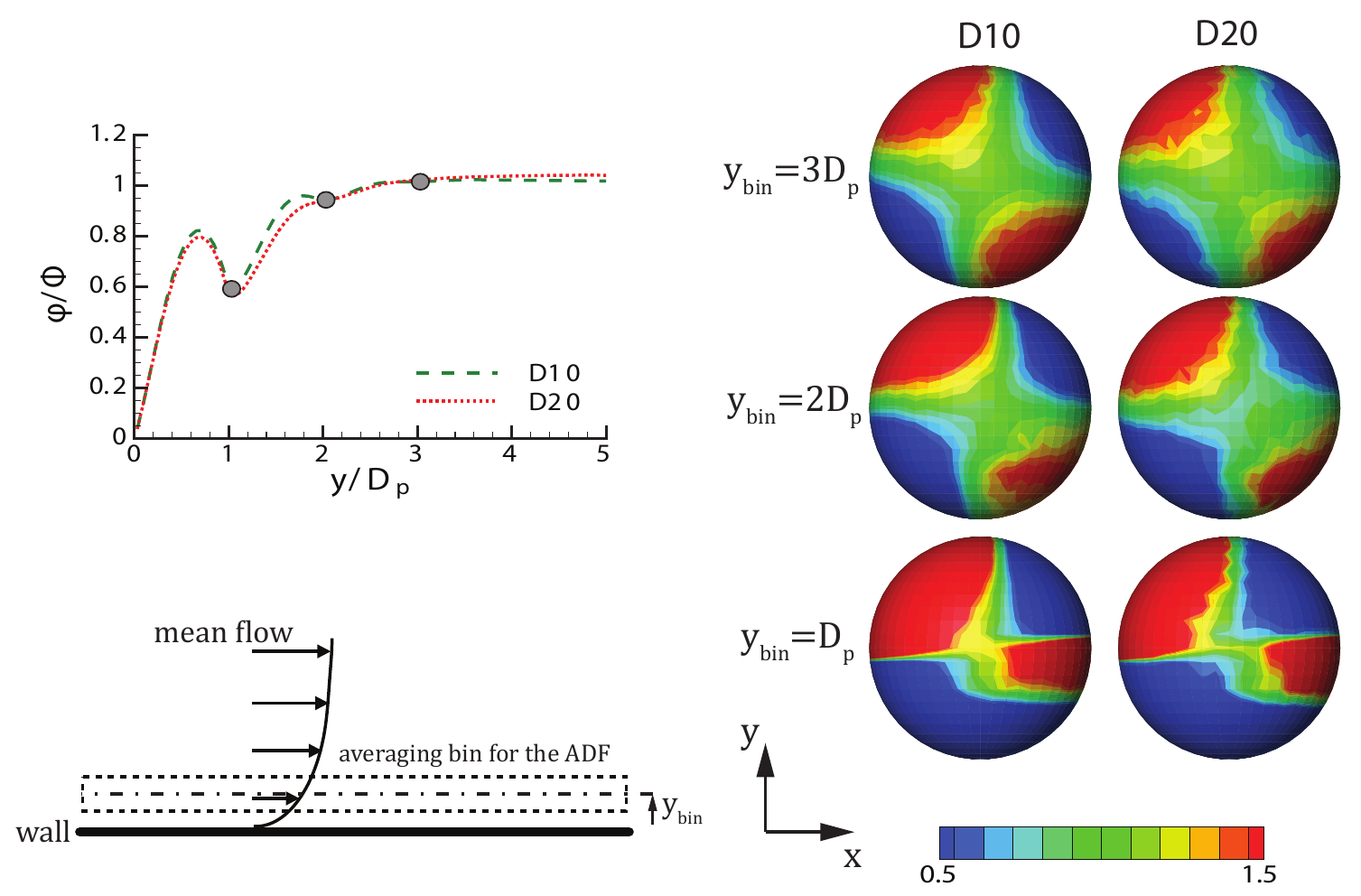}\hfill
\caption{Projection onto the $x-y$ plane of the angular distribution function at contact for simulation D10 (top) and D20 (bottom). Statistics are obtained within wall-normal slots of size $D_p$, centered at $y/D_p=$ $1$, $2$ and, $3$ (flow is from right to left). The curves on the right are the profiles of the mean local volume fraction, $\phi/\Phi$ versus the wall-normal distance $y/D_p$, with the markers denoting the locations where the adf are sampled. \pc{These profiles are uniform for $y/D_p>5$ \citep{Costa-et-al-PRL-2016}. The averaging procedure is illustrated on the bottom-left corner.}}\label{fig:angdist} 
\end{figure}

The particle near-wall layer impacts the scale separation on which the scaling laws for mean velocity are based. It is therefore interesting to understand how the near-wall inhomogeneity affects the particle structure, and how far this inhomogeneity extends in the wall-normal direction. We can quantify this through the angular distribution function at contact (adf). The adf measures the probability of finding a particle pair at a fixed distance $r$, as a function of the polar ($\theta$) and azimuthal ($\varphi$) angles, normalized by the values of a random particle distribution. Formally,
\begin{equation}
g(\theta,\varphi) = \frac{1}{r^2\sin(\theta)}\frac{\mathrm{d}^2 N_{\theta,\varphi}}{\mathrm{d}\theta\mathrm{d}\varphi}\frac{1}{n_0} \,\, \mathrm{;} \,\, n_0 = \frac{N(N-1)}{2V}
\end{equation}
where $N_{\theta,\varphi}$ denotes the number of particles on a segment of a spherical surface of radius $r=D_p/2$, and polar and azimuthal angles within the ranges $\theta'\in[0,\theta]$ and $\varphi'\in[0,\varphi]$. The adf is computed from the simulation data using bins with wall-normal extent $D_p$, centered at $y_{\pc{bin}}/D_p=1$, $2$, and $3$, i.e. at the same wall-normal distance if scaled with the particle size \pc{(see the bottom-left corner of figure~\ref{fig:angdist}).}\par 
The projection of the adf onto the $x-y$ plane is displayed in figure~\ref{fig:angdist} for the two cases under investigation.
For a suspension in homogeneous shear, the particles tend to be preferentially attracted towards each other when located in a compression region (negative local strain) and reciprocally repelled from each other when located in a stretching region \citep{Morris-RA-2009}. This results in a distribution with two planes of symmetry: two regions of higher clustering in the $xy>0$ quadrants and two of lower accumulation in the $xy<0$ quadrants.
The inhomogeneity introduced by the wall, bounding the particle trajectories, induces a deviation from this picture. 
The wall-effect vanishes for wall-normal distances $y/D_p\gtrsim 3$ for both simulations.
Interestingly, both adf agree quantitatively.
As the distance from the wall is normalized by the particle diameter, for fixed volume fraction, the wall inhomogeneity affects the particle dynamics at distances proportional to $D_p$, despite the different local behavior (different wall-normal location in viscous units) of the suspending fluid. \par
\pc{The local volume fraction profile also presented in figure~\ref{fig:angdist} illustrates that the near-wall inhomogeneity vanishes above $y\sim 4D_p$. For higher wall-normal distances this profile is uniform \citep{Costa-et-al-PRL-2016}. It is interesting to note that a denser near-wall layer was existing at lower Reynolds numbers for the same volume fraction \cite[see e.g.][]{Picano-et-al-JFM-2015,Wang-et-al-JFE-2016,Yu-et-al-JoT-2016}. This decrease of layering is attributed to the stronger mixing in this flow due to a higher Reynolds number.}  \par
\begin{figure}
\centering
\includegraphics[width=0.89\textwidth]{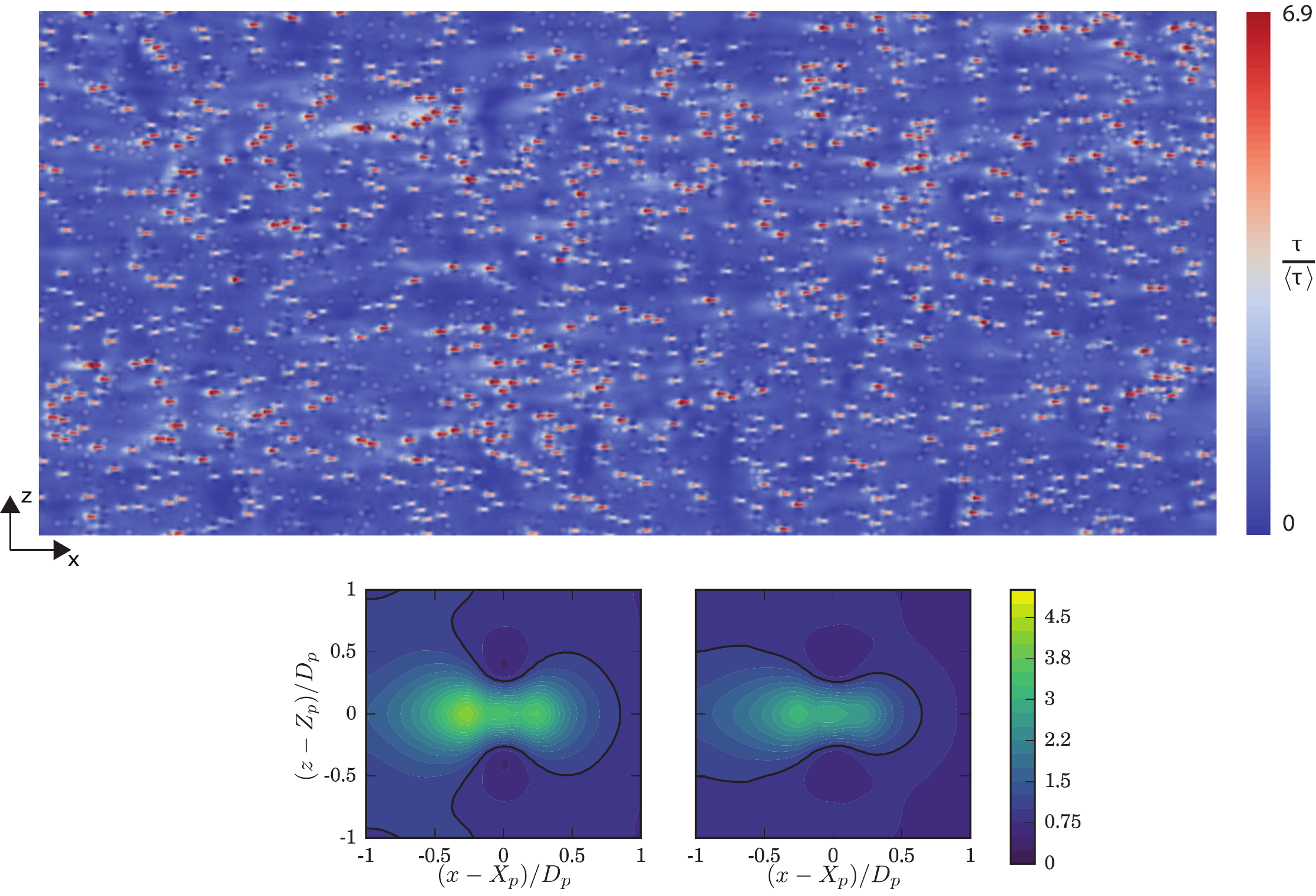}
\caption{Contours of wall shear stresses (normalized with the corresponding mean value) for case D20 (top). Mean wall shear conditioned to the locations of the near-wall particles and normalized with the mean wall shear (bottom) for cases D10 (left) and D20 (right). The data are extracted from the average of the top and bottom walls for one temporal realization.}\label{fig:high_shear_at_particle_positions}
\end{figure}
 Let us now investigate the influence of the particle-wall layer on the distribution of wall shear-stress. To this end, we report in figure~\ref{fig:high_shear_at_particle_positions} contours of the instantaneous wall-shear  for the case D20, together with the particle positions close to the wall (shown with transparency).
The data reveal a strong correlation between regions of high shear and the particle location. 
This can explain the increase in drag with respect to predictions from the continuum limit: high particle-to-fluid apparent slip velocity close to the wall corresponds locally to events of high wall shear stress, which are not present in the `equivalent' single-phase flow with a modified suspension viscosity. Panel (b) of the same figure shows the mean wall shear conditioned to the particle positions. Results for small and large particles show a similar qualitative behavior: 
the shear is higher than the mean around the particles, except close to the surface of the particle, in the spanwise direction. This is a consequence of the reduction of streamwise velocity as the fluid moves around the particle. The contours of the shear stress for the smaller particles show a higher local increase in shear stress, which is, on the other hand, more localized. In fact, the contribution of these \emph{hot spots} of shear stresses to the total shear \pc{(computed from a conditional average of the wall shear stresses in a $2D_p\times 2D_p$ underneath the particle)} is the same for both cases, D10 and D20, and quite high: $69\%$.\par
\begin{figure}
\centering
\includegraphics[width=0.49\textwidth]{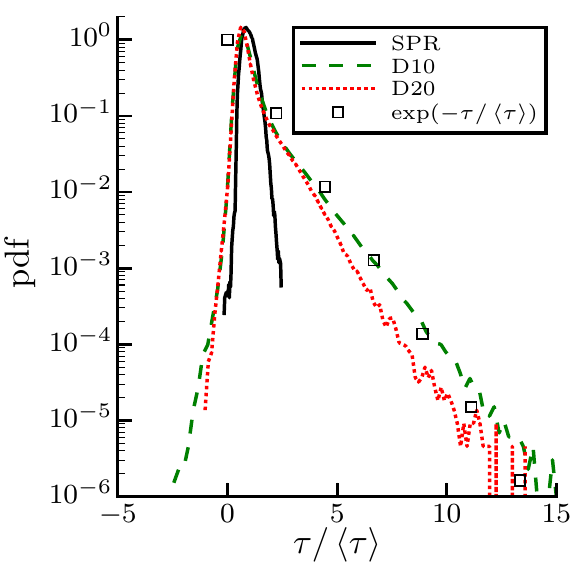}
\includegraphics[width=0.49\textwidth]{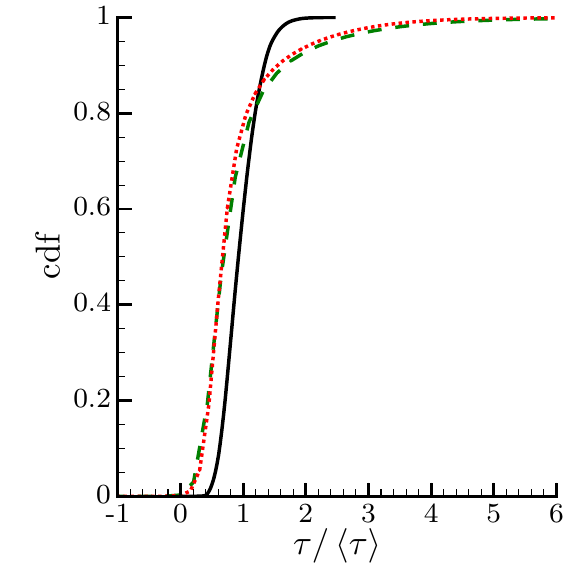}
  \put(-25 ,40 ){\color{black}(b)}
  \put(-190,40 ){\color{black}(a)}
\caption{ (a) Probability density function and (b) corresponding cumulative distribution function of wall shear stress (normalized with its global mean $\rho u_\tau^2$) for cases D20, D10 and SPR (a).}\label{fig:pdfshear}
\end{figure}
The probability density function (pdf) of the shear stress distribution for cases D20, D10 and SPR, presented in figure~\ref{fig:pdfshear}, shows a clear difference between the typical distribution of shear stresses for a single-phase flow, and for the particle-laden cases. In the presence of particles, the tails of the pdf are greatly widened, with very high probability of finding values of the shear stress up to $4$ times the mean value, and likewise for values lower than the mean, including negative events. We note there that
it is reported for single-phase flow that $\tau_w^{+,rms}=\tau_w^{rms}/\left<\tau_w\right>\approx 0.4$ and that the pdf typically follows a log-normal distribution \citep{Orlu-and-Schlatter-PoF-2011}. Indeed, we obtained a value of $\tau_w^{+,rms}=0.39$ for case SPR, with a pdf well-fitted by a log-normal distribution (not shown). The hot-spots of high shear cause a striking difference for the laden cases: $\tau_w^{+,rms}=1.1$ and $1.0$ for cases D10 and D20. Moreover, the pdf no longer fits a log-normal distribution. The hot-spots of high wall shear induce a \textit{kink} in the pdf for $\tau \gtrsim \left<\tau\right>$, resulting in a clear exponential tail, fitted by $\lambda\exp(-\lambda\tau/\left<\tau\right>)$ with $\lambda \approx 1$. Further, case D10 shows higher probability of extreme events, (which, again, are more localized). The largest values attained by the shear stress that in single-phase flows do not contribute significantly to the mean shear, account for about $10\%$ of the total in particle-laden channel flows, as shown by the cumulative probability distribution function (cdf) displayed in panel (b).\par
\begin{figure}
\centering
\includegraphics[width=0.99\textwidth]{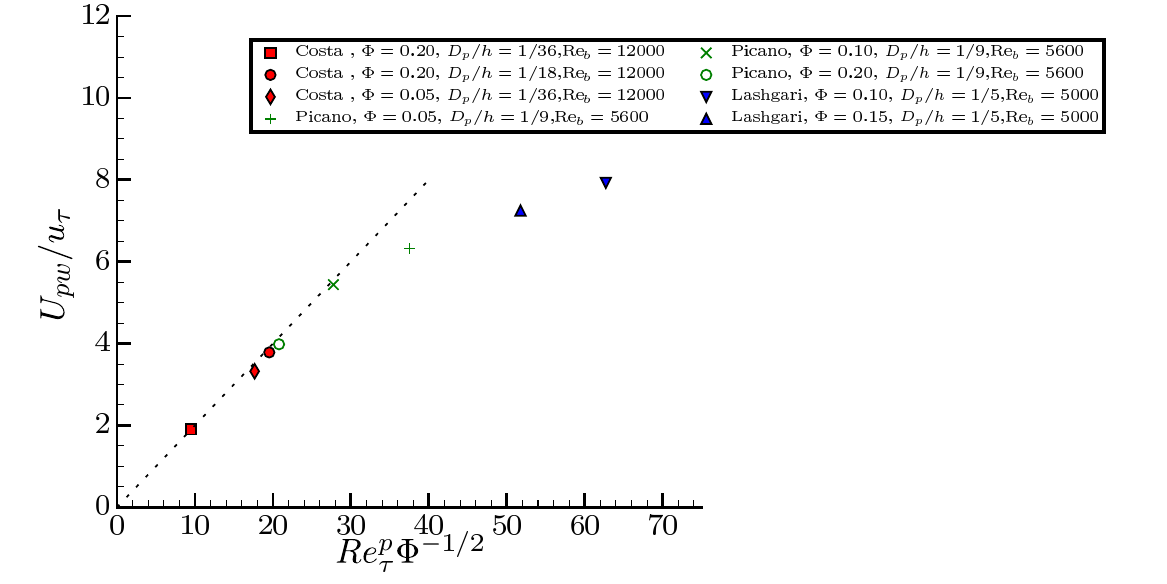}
\caption{Particle-to-fluid slip velocity $U_{pw}$ normalized with the wall friction velocity $u_\tau$ vs $Re_\tau^p\Phi^{-1/2}$ (see definition in the text). To support the validity of the proposed scaling, data from \cite{Picano-et-al-JFM-2015} and \cite{Lashgari-et-al-PRL-2014} are also included.}\label{fig:utau_vs_upw}
\end{figure}

The results presented so far shed light on the relation between the wall slip velocity $U_{pw}=u_p(y=0)$ and the wall friction $u_\tau$ (recall figure~\ref{fig:mean_vel} b). Assuming that (1) the average shear in the hot-spots matches the average wall shear and (2) the particles impose a mean shear that scales with $U_{pw}/D_p$ over an area $D_p L_w$ ($L_w$ being the streamwise extent of the hot-spot, which as seen in figure~\ref{fig:high_shear_at_particle_positions} is significantly larger than $D_p$), we can write:
\begin{equation}
N_{pw} D_p L_w \mu \frac{U_{pw}}{D_p} = \rho u_\tau^2 L_x L_z\mathrm{.}
\end{equation}
with \pc{$N_{pw}$} the number of particles in this layer. Further noting that $N_{pw}D_p^3/(D_pL_xL_z)\sim\Phi$, we have 
\begin{equation}
\Phi\frac{L_w}{D_p}\frac{U_{pw}}{D_p} \sim \frac{u_\tau^2}{\nu}\mathrm{,}
\end{equation}
and finally, assuming that $L_w/D_p$ scales as the mean-free-path in a two-dimensional layer of particles, i.e., $\propto \Phi^{-1/2}$, we obtain 
\begin{equation}
\frac{U_{pw}}{u_\tau} = C_{pw} Re_\tau^{p} \Phi^{-1/2}\mathrm{,}
\end{equation}
with $Re_\tau^{p}=u_\tau D_p/\nu$ a particle friction Reynolds number. The data in figure~\ref{fig:utau_vs_upw}, including previous cases in literature, confirm the validity of this relation with  $C_{pw}\approx 0.2$. The data points that do not follow the predicted scaling 
correspond to very large particles ($h/D_p = 5$), for which the influence of the outer scales on the near-wall particle dynamics compromises the first assumption above, and to low volume fractions.\par
\subsection*{Homogeneous suspension region (HSR)}
\begin{figure}
\centering
\includegraphics[width=0.49\textwidth]{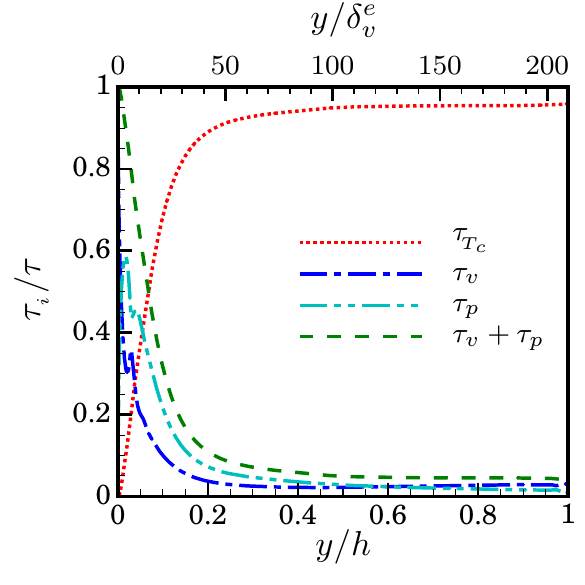}\hfill
\includegraphics[width=0.49\textwidth]{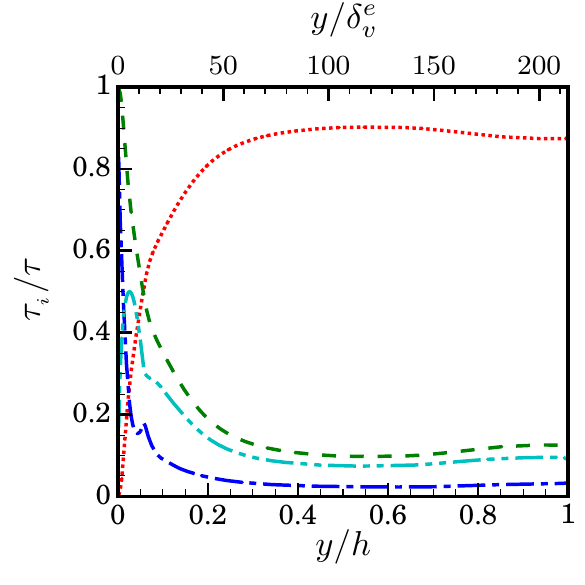}
  \put(-45 ,50){\color{black}(b)}%
  \put(-230,50){\color{black}(a)}%
  \\
\includegraphics[width=0.49\textwidth]{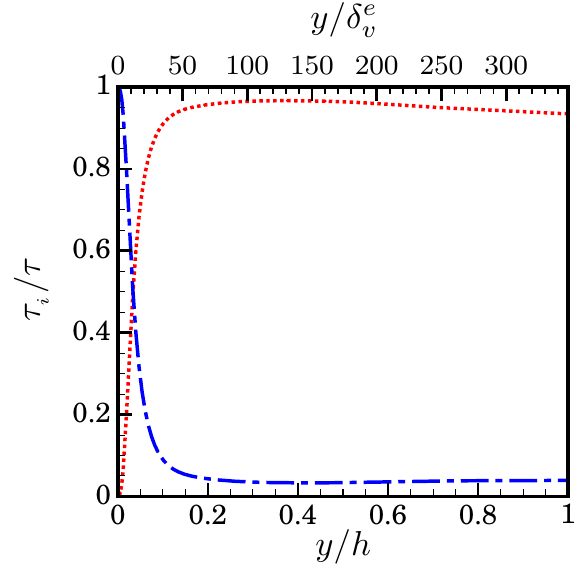}\hfill
\includegraphics[width=0.49\textwidth]{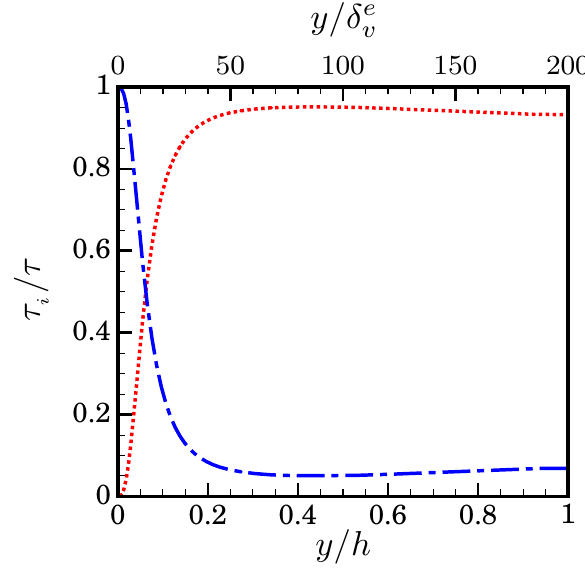}
  \put(-45 ,50){\color{black}(d)}%
  \put(-230,50){\color{black}(c)}%
\caption{Stress contributions $\tau_i(y)/(u_\tau^2(1-y/h))$ for the different cases D10 (a), D20 (b), SPR (c) and CLR (d), see equation~\eqref{eqn:stress_budget}. $\tau_{T_c}=\tau_{T_f}+\tau_{T_p}$ denotes the contribution of the Reynolds stresses of the combined phase.}\label{fig:stress_contribution}
\end{figure}

%The remainder of the paper discusses fluid and particle dynamics in the different domain regions, with focus on the dynamics of the homogeneous suspension region and wall-normal propagation of the inhomogeneity caused by the particle-wall layer. 
The contribution of the different stresses to the suspension streamwise momentum transport (i.e. the stress budget) can be derived from volume- and ensemble-averaging the mixture streamwise momentum equation \cite[see][]{Marchioro-et-al-IJMF-1999,Picano-et-al-JFM-2015}. It reads for plane channel flow 
\begin{equation}
  \left<\tau\right> = u_\tau^2\left(1-\frac{y}{h}\right) = \underbrace{(1-\phi)\left<- u_f'v_f'\right>}_{\tau_{{T}_f}} + \underbrace{(1-\phi) \nu \frac{\mathrm{d}\,U_f}{\mathrm{d}\,y}}_{\tau_v} + \underbrace{\phi\left< -u_p'v_p'\right>}_{\tau_{{T}_p}} + \tau_p\mathrm{,}\label{eqn:stress_budget}
\end{equation}
where the terms on the right-hand-side denote the fluid Reynolds stresses $\tau_{{T}_f}$, the viscous stresses $\tau_v$, the particle Reynolds stresses $\tau_{{T}_p}$, and the particle stresses $\tau_p$, the latter related to the stresslet, moment of the material acceleration acting on a particle and inter-particle collisions \citep{Guazzelli-and-Morris-2011}; in the expression above\pc{,} $\phi$ denotes the mean local solid volume fraction.
The wall-normal profiles of the different contributions to the total stress (obtained dividing equation~\eqref{eqn:stress_budget} by $u_\tau^2(1-y/h)$) are reported in figure~\ref{fig:stress_contribution}.
As also observed in \cite{Lashgari-et-al-IJMF-2016,Lashgari-et-al-PRL-2014,Picano-et-al-JFM-2015} the profiles of particle stresses reach a maximum at $y = D_p/2$, at the location of the local maximum of the profiles of the local volume fraction (see figure~\ref{fig:angdist}), followed by a minimum at $y\sim 2D_p$. The Reynolds stresses are relatively small in this region, and therefore the local maximum/minimum of the particle stresses is compensated by a local minimum/maximum of the viscous shear stresses. This is caused by the strong shear that the first layer of particles, flowing with significant slip velocity, imposes on the fluid above it. Further away from the wall, the profiles follow a trend that resembles the one of the shear-stress profile, suggesting a linear (Newtonian) scaling of the particle stresses with the shear rate.\par

The deviation from the continuum limit (CLR) can be examined by comparing the sum of the particle and viscous stresses to the profile of viscous stresses in panel (d) of figure~\ref{fig:stress_contribution} (note that an effective viscosity incorporates the effects of both viscous and particle stresses). As expected, the case with smaller particles is much closer to this continuum limit than the case with larger particles \citep[see also][]{Costa-et-al-PRL-2016}. For the latter, while the profile pertaining to the viscous contribution in the bulk is still close to the one of case D10, the particle stress contribution is much larger. This difference can be attributed to the \emph{inertial-shear-thickening} mechanism due to excluded volume effects proposed in \cite{Picano-et-al-PRL-2013}, as the particle Reynolds number based on the particle slip velocity for case D20 is much larger (recall Figure~\ref{fig:mean_vel}~(b)). We should stress here the importance of defining the viscous wall unit from the effective suspension viscosity: if the profiles of cases D10 and D20 would be compared to CLR using the classical inner scaling (where the fluid viscosity and wall friction velocity are used to scale the wall-normal distance) these would not match, as the abscissa on the upper axis $y/\delta_v^e$ would have to be multiplied by $\nu^e/\nu = 1.89$ in panels (a) and (b).\par
\begin{figure}
\centering
\includegraphics[width=0.36\textwidth]{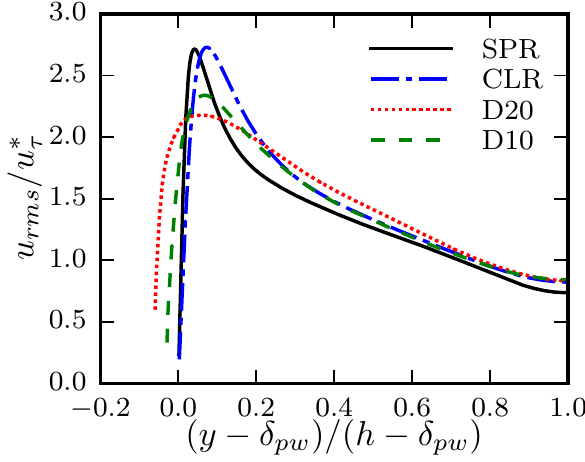}\hspace*{5mm}
\includegraphics[width=0.36\textwidth]{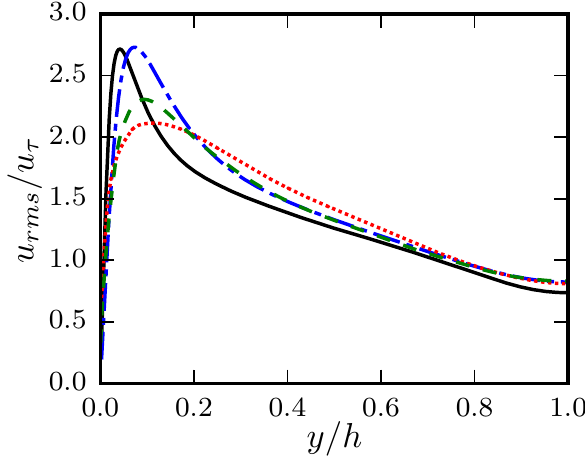}
  \put(-25 ,25){\color{black}(b)}%
  \put(-180,25){\color{black}(a)}%
  \\
\includegraphics[width=0.36\textwidth]{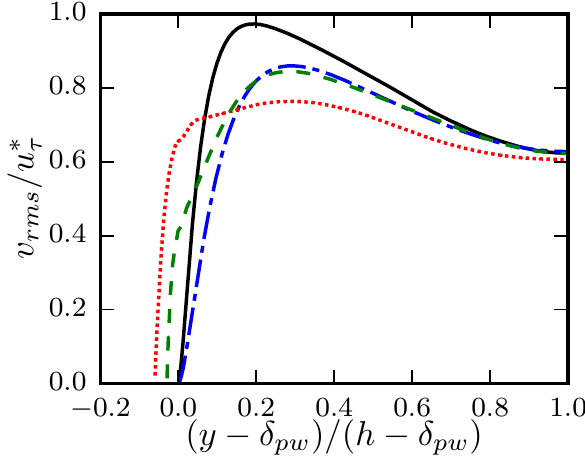}\hspace*{5mm}
\includegraphics[width=0.36\textwidth]{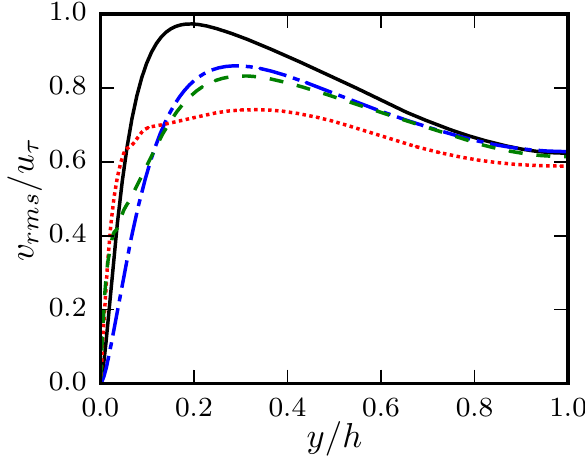}
  \put(-25 ,25){\color{black}(d)}%
  \put(-180,25){\color{black}(c)}%
  \\
\includegraphics[width=0.36\textwidth]{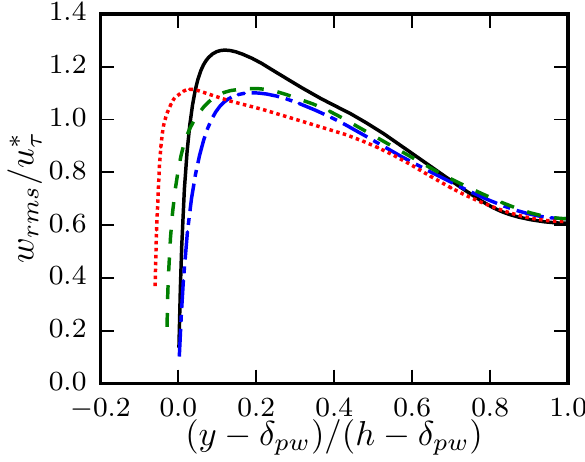}\hspace*{5mm}
\includegraphics[width=0.36\textwidth]{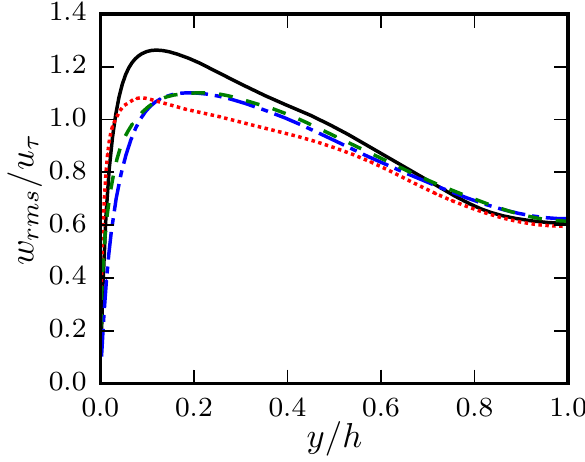}
  \put(-25 ,25){\color{black}(f)}%
  \put(-180,25){\color{black}(e)}%
  \\
\includegraphics[width=0.36\textwidth]{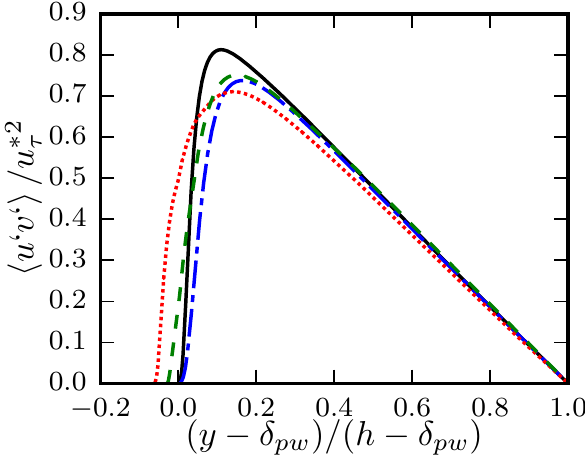}\hspace*{5mm}
\includegraphics[width=0.36\textwidth]{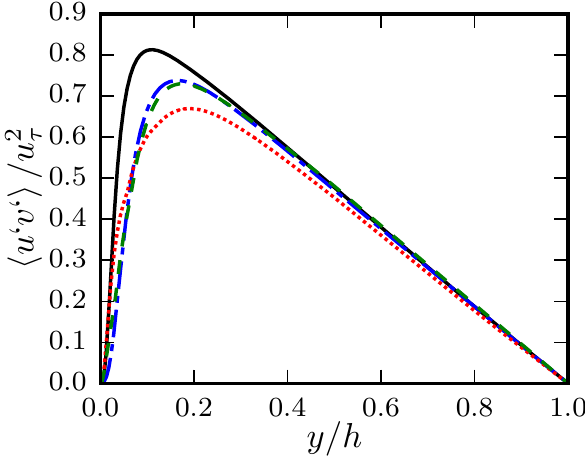}
  \put(-25 ,45){\color{black}(h)}%
  \put(-180,45){\color{black}(g)}%
\caption{Profile of rms of fluctuating streamwise, spanwise and wall-normal fluid velocity from the different simulations. Classical scaling is used in panels (a,c,e,g), whereas the scaling laws of \cite{Costa-et-al-PRL-2016} are used in panels (b,d,f,h).}\label{fig:rms_fluid_inn}
\end{figure}

As for the other quantities presented so far, we expect the second-order Eulerian statistics for the laden cases to approach those of CLR with decreasing particle size. These are shown in figure~\ref{fig:rms_fluid_inn} where we depict the profiles of the root-mean-square (rms) of the fluctuating fluid velocity and Reynolds stresses from the different simulations. When comparing to the single-phase flow statistics we also observe, as reported in  \cite{Picano-et-al-JFM-2015}, that the fluctuating fluid velocity close to the wall is enhanced by the presence of the particles.\par
Let us take one step further and test the scaling arguments of \citet{Costa-et-al-PRL-2016} for the second-order statistics. These can be compared directly to the case CLR, as it is meant to be the same flow, finite size effects aside. We therefore correct the wall-normal distance for the presence of the particle-wall layer through a virtual wall origin, $\delta_{pw}$, and correct velocity scale for inner-scaling, $u_\tau^*=u_\tau(1-\delta_{pw}/h)^{1/2}$ (panels a,c,e and g in the figure). 
The magnitude of near-wall peaks in streamwise velocity rms, typically located at $y=12\delta_v\lesssim D_p$, are not recovered in the interface-resolved simulations. Despite this, away from the wall (where the dynamics is only weakly affected by the particle-wall layer due to sufficient scale separation) the profiles show good agreement. Particularly, the profiles pertaining case D10 show a remarkable collapse over the entire outer region for all three components of the velocity rms and Reynolds stresses, even without correction. The same cannot be safely stated for the case with larger particles (D20) where finite size effects, also present away from the particle-wall layer, have an influence on the statistics in the bulk of the flow. Despite these small deviations for case D20, it is clear that, away from the wall, the second-order statistics are described with good approximation by those of a Newtonian fluid with an effective suspension viscosity.\par
\subsection*{Particle Dynamics}
\begin{figure}
\centering
\includegraphics[width=1.0\textwidth]{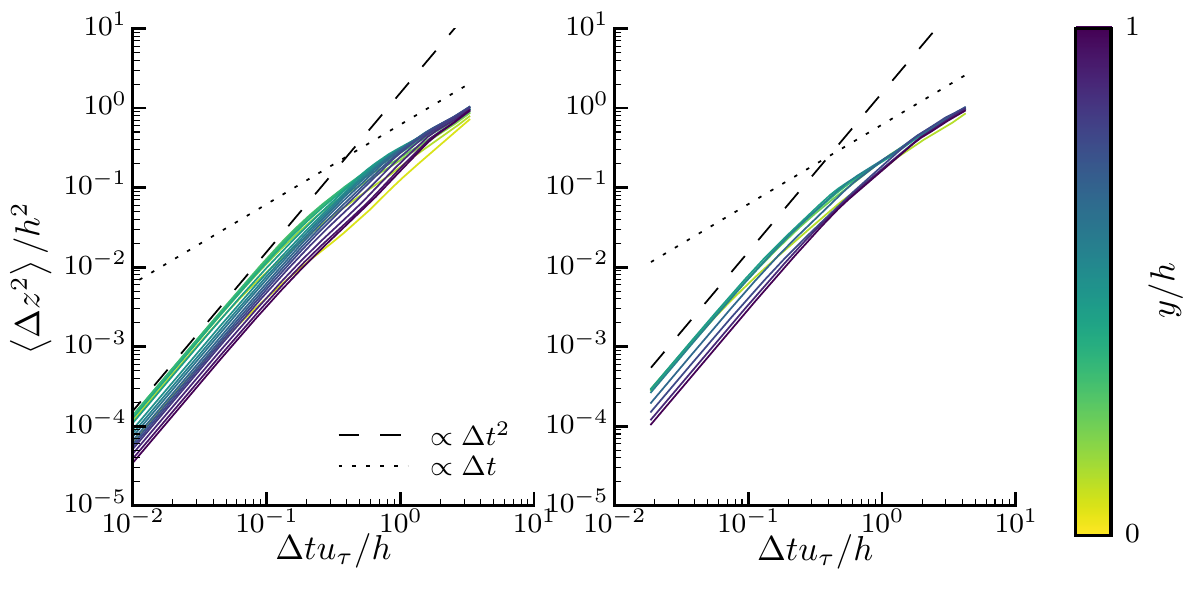}
  \put(-80 ,75){\color{black}(b)}%
  \put(-240,75){\color{black}(a)}%
\caption{Single-particle mean square spanwise displacement $\left<\Delta z_1^2\right>(\Delta t,y)$ normalized with the half channel height $h^2$, for cases  (a) D10 and (b)  D20 versus time in units of the integral time scale $h/u_\tau$. The colors indicate different wall-normal distances.}\label{fig:msqdsp}
\end{figure}

\pc{Analysing the mean flow solely in an Eulerian framework does not give direct insights on the dynamics of individual particles. To \pc{better} understand the particle dynamics, we further explore the DNS dataset by computing Lagrangian statistics}. We first focus on the evolution of the single-point mean-square displacement of particles in the spanwise direction 
to prevent biases due to statistical inhomogeneity in the wall-normal, and non-zero mean flow in the streamwise direction \citep[see also][]{Lashgari-et-al-IJMF-2016}. This is defined as
\begin{equation}
\left<\Delta z_1^2\right>(\Delta t,y) = \left<\left(z(t+\Delta t)-z(t)\right)^2\right> \label{eqn:msqdisp_sp}
\end{equation}
where $\left<\right>$ denotes the time average over all particles located in a bin centered at a wall-normal position $y$ at time $t$ and wall-normal extent $D_p$. 
This observable is displayed in figure~\ref{fig:msqdsp} versus time in units of a characteristic integral scale of the turbulent fluid motion, $h/u_\tau$.
One can \pc{distinguish} two well-known regimes: the ballistic regime where the high temporal correlation results in a mean square displacement $\left<\Delta z_1^2\right>\propto \Delta t^2$ and the diffusive regime where the motion decorrelates from the initial sampling instant and $\left<\Delta z_1 ^2\right>\propto \Delta t$ \cite[see e.g.][]{Sierou-and-Brady-JFM-2004}.
In the ballistic regime, particles reach the fastest dispersion at the distance from the wall corresponding to the peak in particle velocity rms (see figure~\ref{fig:rms_fluid_inn}).
The diffusion coefficient (slope of the profile in the diffusive regime) is nearly the same for the both cases, away from the wall. Two main mechanisms are responsible for the spanwise particle displacements and the subsequent self-diffusion: (1) short-range inter-particle interactions and (2) interactions with turbulent structures of dimension larger than the particle size. 
One can picture a sequence of these events as successive random-walk steps during the particle motion. Indeed, \cite{Lashgari-et-al-IJMF-2016} observed 
in the turbulent regime an increase of the diffusion coefficient with increasing Reynolds number for fixed volume fraction, and a milder increase with increasing volume fraction at fixed Reynolds number. 
The smaller diffusion coefficients observed near the wall can therefore be explained by the constraint that the wall confinement imposes on the particle motion, and by the reduction of the characteristic turbulent integral scale, estimated as $l_m(y)\sim \min(\kappa y,0.1h)$ \pc{note: I changed to $l_m(y)$}, with $l_m$ being a mixing length \cite[][]{Pope-2001}. The turbulent length scales are hindered up to a point where $l_m \sim D_p$, the value below which turbulent fluctuations have a weak influence on the particle kinematics. Despite the relatively small particle inertia, the temporal filtering due to the higher response time of the larger particles \citep{Hinze-1975} can also cause a decrease in diffusion coefficient. We expect however the latter effect to be milder, as increasing particle inertia through an increased mass density (while fixing the other governing parameters) has a small influence on spanwise particle dispersion, as shown in \cite{Fornari-et-al-PoF-2016}.\par
The diffusion coefficient is larger for the flow with smaller particles (D10), even though the collision probability is larger for the largest particles (as suggested by the asymptotic limit of vanishing particle size at fixed volume fraction, and will be confirmed later). We can therefore conclude that, at the volume fraction under consideration here, the turbulent fluid motion is the main source of self-diffusion and, consequently, larger particles result in slower dispersion. Note also that the diffusive regime is reached away from the wall at times $\Delta t = O(h/u_\tau)$ of the same order as the turnover time of large eddies in the bulk, consistently with this observation.\par
\begin{figure}
\centering
\includegraphics[width=0.99\textwidth]{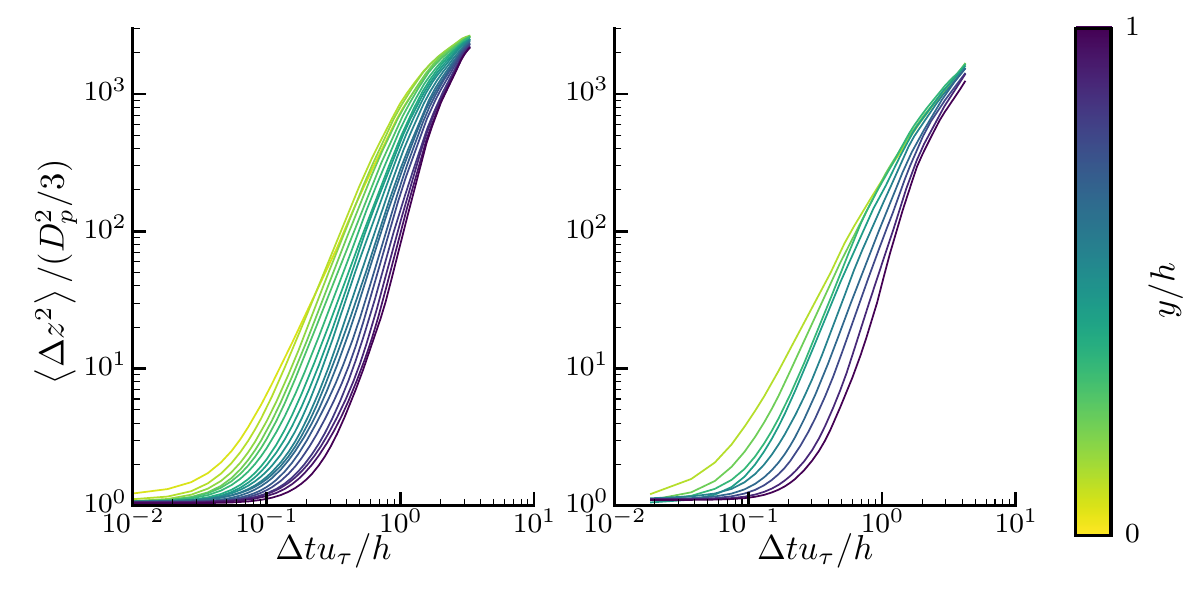}
  \put(-80 ,70){\color{black}(b)}%
  \put(-240,70){\color{black}(a)}%
  \\
\includegraphics[width=0.99\textwidth]{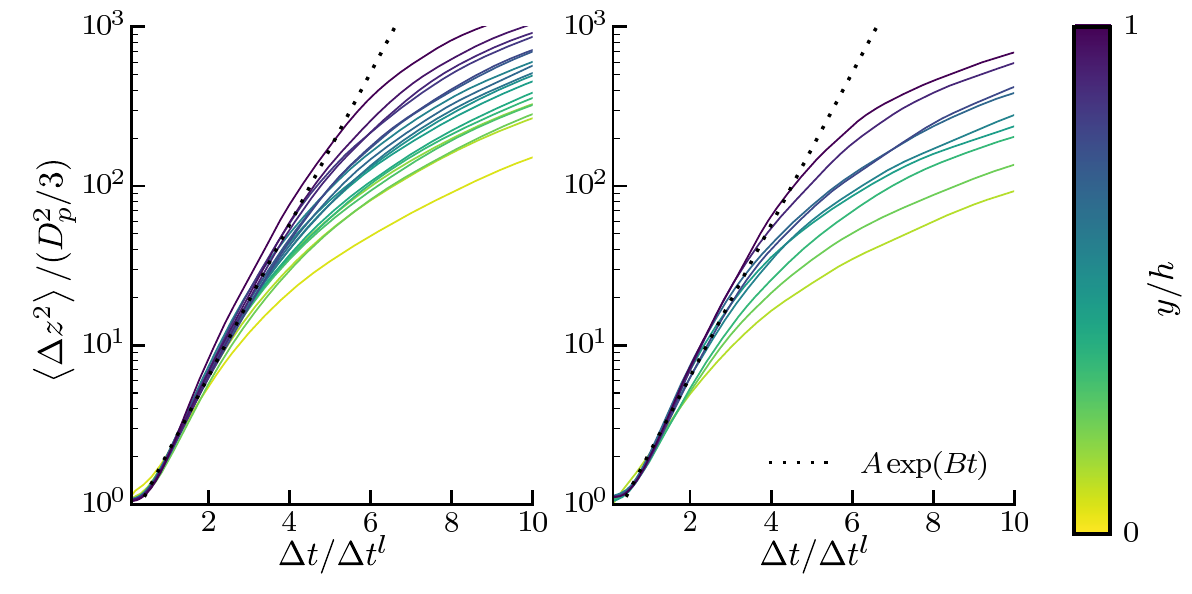}
  \put(-80 ,70){\color{black}(d)}%
  \put(-240,70){\color{black}(c)}%
\caption{Two-particle mean square spanwise displacement $\left<\Delta z_2^2\right>(\Delta t,y)$ normalized with the mean initial spanwise square distance at contact $D_p^2/3$ for  (a) case D10 and  (b) case D20 versus time normalized with by the integral time scale $h/u_\tau$.
The different lines are color-coded to indicate the particle wall-normal distance. Panels (c) and (d) illustrate the initial exponential scaling by plotting the same quantity versus $\Delta t^l=\Delta t|_{<\Delta z_2^2>=2D_p^2/3}$ (see discussion in the text). 
The dotted-lines indicate the fitting function $A\exp(Bt)$ with $A=0.7$ and $B=1.1$.}\label{fig:msqdsp_2p}
\end{figure}

Further insight into the particle dynamics can be gained by examining the  pair-dispersion statistics. 
The two-point mean spanwise square displacement is displayed
in figure~\ref{fig:msqdsp_2p} 
for two particles at contact at $t=t_0$,
\begin{equation}
\left<\Delta z_2^2\right>(\Delta t,y) = \left< \left(\delta z(t_0+\Delta t)-\delta z(t_0)\right)^2\right>\mathrm{,}
\label{eqn:msqdisp_2p}
\end{equation}
where $\delta z$ denotes the spanwise interparticle distance and $t_0$ the instant at which the particles are in contact.
For short time scales, lubrication dominates and the particles display a highly correlated motion. The duration of this regime increases with the distance to the wall. This slower relative dispersion is linked to the decreasing relative inter-particle velocity towards contact with increasing wall-normal distance, to be quantified later.
The large particles, case D20, show faster pair dispersion, also due to higher relative inter-particle velocity towards contact: it turns out, as elaborated later when the collisional dynamics is addressed, that this quantity scales with the particle Reynolds number.
At later times, the particle dispersion $\left<\Delta z_2^2\right>\propto \Delta t^\alpha$, with $2<\alpha<3$. Note that for tracer particles in a homogeneous isotropic turbulent flow $\alpha = 3$ for separation distances within the inertial subrange \citep{Salazar2009}. For larger separation distances the diffusive limit is recovered.\par
 The mean square displacement of tracer particles in turbulence with sub-Kolmogorov separation distance (i.e.\ in the \emph{dissipation subrange})  grows exponentially in time as the relative particle velocity and the separation are proportional \citep{Batchelor1952,Salazar2009}: i.e.\ $\left<\delta z_2^2\right>=\left<\delta z_0^2\right>\mathrm{e}^{2\Delta t/t_c}$, where $t_c$ is a characteristic inter-particle response time. Although in the present study $D_p$ is $O(10)$ times larger than the smallest turbulence scale, we observe a clear exponential growth at short times, just after the first highly correlated regime when $\left<\Delta z_2^2\right>$ is approximately constant. 
 The time at which this exponential regime sets in corresponds to a mean square separation of about ${2D_p^2/3}$. This exponential regime is highlighted in the bottom panels of figure~\ref{fig:msqdsp_2p} where the time on the horizontal axis is divided by  $\Delta t^l=\Delta t|_{<\Delta z_2^2>=2D_p^2/3}$. 
 This growth is due to a different mechanism than the one of tracer particles, possibly the combination of a uniform shear force (note that under uniform shear $\delta z$ is proportional to the relative velocity) following a short-range lubrication interaction.\par
Finally, we investigate the particle collisional dynamics. A collision event takes place when (1) the particles are at sufficiently close distance and (2) their relative velocity drives them towards contact. These two factors are investigated separately. The first by the radial distribution function (rdf), which quantifies the probability of finding a second particle at distance $r$ normalized by the probability of a random distribution of particles:
\begin{equation}
  g(r,y) = \frac{1}{4\pi r^2}\frac{\mathrm{d} N_r}{\mathrm{d}r}\frac{1}{n_0}\mathrm{,}
\end{equation}
where $N_r$ denotes the number of particle pairs in a spherical volume of radius $r$. Thus, if $g(r,y)$ assumes values larger than  $ 1$, particles are preferentially sampled. The second observable is the distribution of %incoming (i.e.\ towards contact) 
the relative particle velocity projected in the direction of the line-of-centers $\Delta v^{n,-}$, given for two particles $i$, $j$ with velocities $\mathbf{u}_{i/j}$ and positions $\mathbf{x}_{i/j}$ by:
\begin{equation}
  \Delta v^{n,-}(r,y) = \max\left\{0,-(\mathbf{u}_j-\mathbf{u}_i)\cdot\frac{\mathbf{x}_j-\mathbf{x}_i}{|\mathbf{x}_j-\mathbf{x}_i|}\right\}\mathrm{;}\label{eqn:deltavn}
\end{equation}
where the $\max\{\}$ operator samples the relative velocity \emph{towards} contact, as the superscript `$-$' suggests.
The product of these quantities measures the rate at which particles approach each other. Its value for $r=D_p$ is the so-called collision kernel, $\kappa_c$, the probability of a collision event.\par
\begin{figure}
\centering
\includegraphics[width=0.30\textwidth]{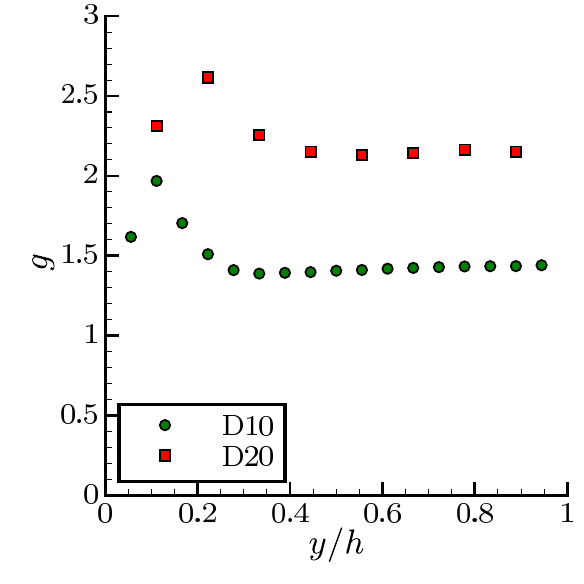}
\includegraphics[width=0.30\textwidth]{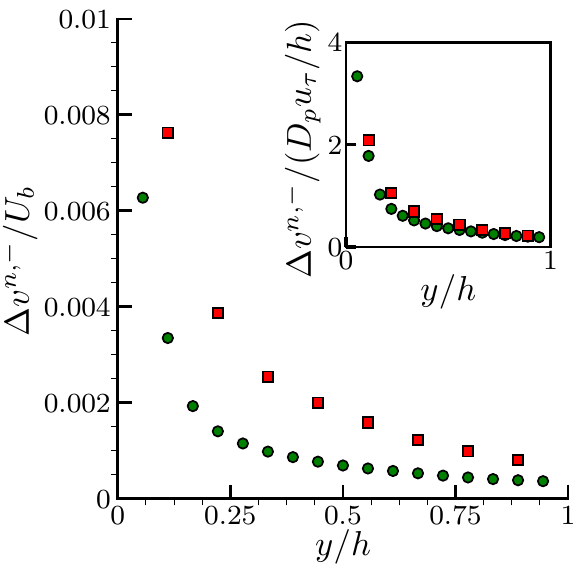}
\includegraphics[width=0.30\textwidth]{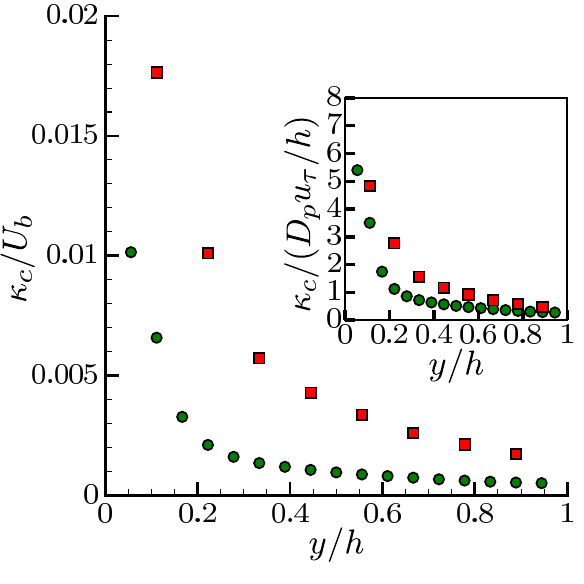}
\put(-260,30){\color{black}(a)}
\put(-140,30){\color{black}(b)}
\put(- 20,30){\color{black}(c)}
\caption{(a) radial distribution function $g(D_p)$ at contact; (b)
normal relative velocity at contact $\Delta v^{n,-}(D_p)/U_b$ and (c) collision kernel $\kappa_c/U_b$ versus the outer-scaled wall-normal distance $y/h$. The insets of panels (b) and (c) show the same quantity scaled with the velocity scale $u_\tau D_p/h$, see equation~\eqref{eqn:scaling}.}\label{fig:contact_stats}
\end{figure}

As for the dispersion statistics, we take into account the inhomogeneity in the wall-normal direction by averaging in wall-parallel bins with wall-normal extent $2D_p$. Figure~\ref{fig:contact_stats} presents the radial distribution function, negative particle relative velocity and the approach rate at contact $r=D_p$.
The profile of $\Delta v^{n,-}$ shows that the dominant mechanism driving particles towards each other is shear from particles at different wall-normal locations. Note that the shear rate away from the PWL is similar for both simulations considered (see figure~\ref{fig:stress_contribution} and table~\ref{tbl:frictionre}). Assuming that particles are driven towards each other by shearing a layered arrangement of particles:
\begin{equation}\label{eqn:scaling}
\Delta v^{n,-} \propto D_p \frac{\partial u}{\partial y} \sim D_p \frac{u_\tau}{h}\mathrm{,}
\end{equation}
with $u_\tau/h$ an estimate of the shear rate in the bulk of the flow. This scaling is tested in the inset of panel (b) and (c), yielding a  better agreement of the profiles of $\Delta v^{n,-}$ and  $\kappa_c$, despite the differences in $g(D_p)$. Note that $u_\tau/h$ is approximately the same for both cases (table~\ref{tbl:frictionre}). The higher values of $g$ at close distances for case D20 are also compatible with this picture, as larger particles are more prone to be driven towards each other by a shear-induced relative motion.\par
\pc{The mechanism governing energy dissipation during particle collisions can be predicted from the impact Stokes number, which can be computed directly from the mean relative approach velocity close to contact: $\mathrm{St}=(1/9)\rho_p\Delta v^{n,-}_{max} D_p/\mu$. The Stokes number corresponding to the maximum value of $\Delta v^{n,-}$, shown in figure~\ref{fig:contact_stats}(b) (case D20), is $\mathrm{St} \approx 0.28$. This average Stokes number is therefore much lower than the critical value of $10$ everywhere in the domain, suggesting that the collisional dynamics are governed by short-range hydrodynamic interactions, with strong viscous dissipation and no elastic rebound; see \cite{Legendre-et-al-CES-2006} and \cite{Yang-and-Hunt-PoF-2006}.}\par
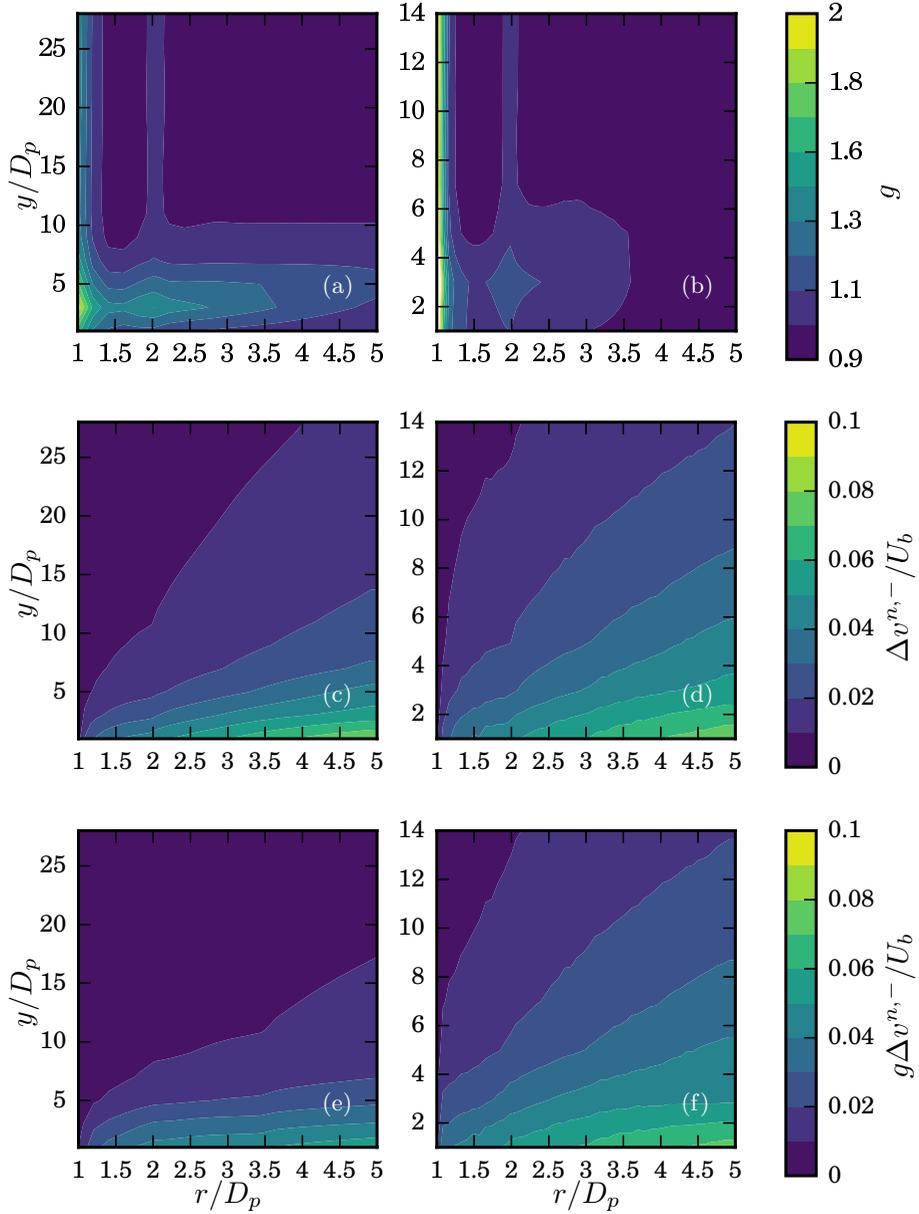
\begin{figure}
\centering
\includegraphics[width=0.99\textwidth]{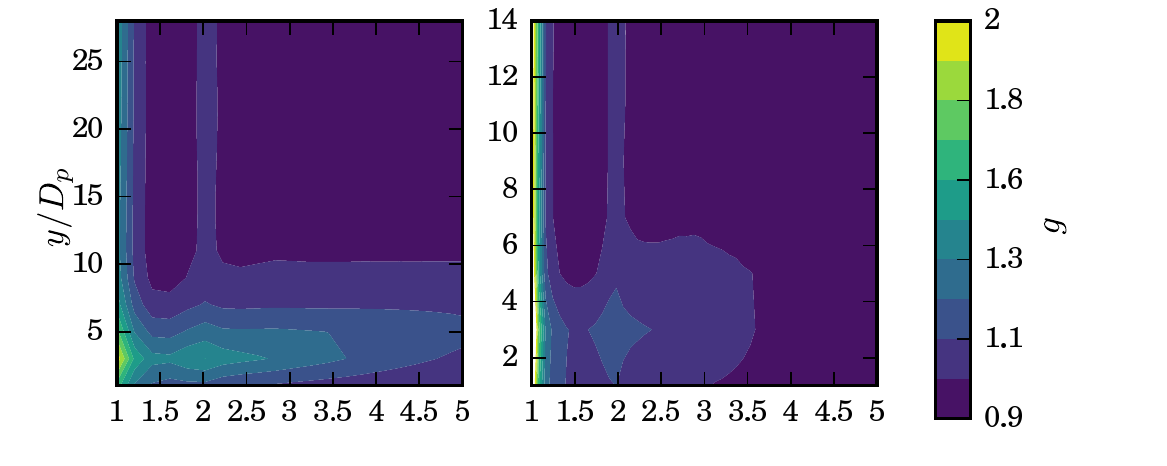}
\put(-250,40){\color{white}(a)}
\put(-115,40){\color{white}(b)}
\\
\includegraphics[width=0.99\textwidth]{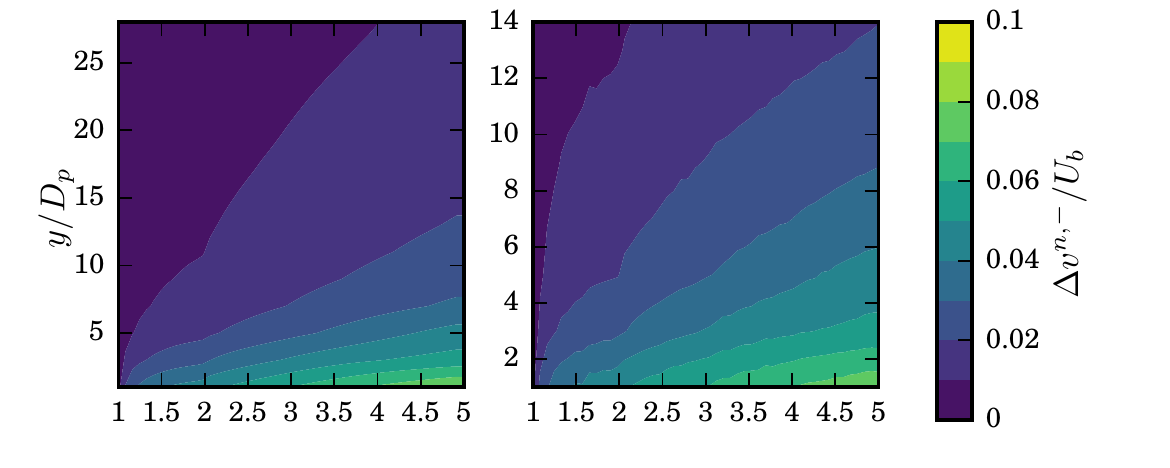}
\put(-250,40){\color{white}(c)}
\put(-115,40){\color{white}(d)}
\\
\includegraphics[width=0.99\textwidth]{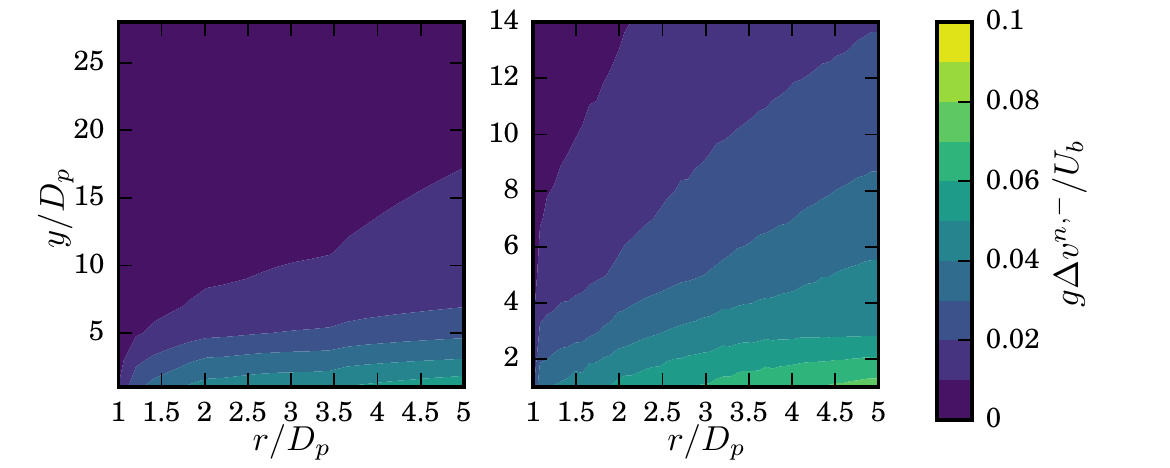}
\put(-250,40){\color{white}(e)}
\put(-115,40){\color{white}(f)}
\caption{Contour plot of radial distribution function, $g$ (top), negative relative velocity $\Delta v^{n,-}/U_b$ (middle), and rate-of-approach $g\Delta v^{n,-}/U_b$ (bottom) for simulation D10 (panels on the left: a,c,e) and D20 (panels b,d,f) as a function of the wall-normal coordinate ($y$) and separation ($r$) scaled with the particle diameter $D_p$.}\label{fig:collisions}
\end{figure}

Figure~\ref{fig:collisions} shows contour plots of the same quantities as in figure \ref{fig:contact_stats}, now as a function of both the separation distance $r$ and the wall-normal coordinate $y$. 
The contours of $g$ show that this quantity is weakly dependent on the wall-normal coordinate, with local maxima at $r/D_p\approx 1$ and $2$, consistent with the presence of statistically significant particle pairs and triplets. It is also interesting to notice that the global maximum of $g$ occurs at a finite distance to the wall. 
Comparing the two simulations, the maximum is located at the same wall-normal distance when scaled with the particle diameter $y\sim 3 D_p$, in agreement with the observation above  that wall-confinement effects are noticeable at distances proportional to the particle diameter (see figure~\ref{fig:angdist}).
The maxima of $g(r)$ correspond therefore to  the optimal trade-off between high shear (driving particles at different wall-normal locations towards each other) and low confinement; indeed further away from the wall, where the mean shear is relatively low,  $g$ becomes almost independent of $y$. 
The maxima of $g$ close to the wall also suggest that larger particles have higher probability of forming particle pairs, and lower probability of forming triplets.\par
Panels (c) and (d) of the same figure display contours of $\Delta v^{n,-}$. For fixed separation distance $r$, the average inter-particle approaching velocity decreases with the wall-normal distance \pc{$y$}, which can be explained by a decreasing local shear rate: the lower the shear, the smaller shear-induced differences of the particle relative velocities. \pc{Consistently,} the variations of $\Delta v^{n,-}$ with $y$ are larger close to the wall. The negative relative particle velocity is larger for the largest particles, case D20, over the entire flow, \pc{as the scaling suggested in equation~\eqref{eqn:scaling} predicts}. Finally, we report the rate-of-approach $\Delta v^{n,-}g$ in panels (e) and (f). Clearly, the differences in $g$ discussed above induce significant differences in this quantity only close to contact. At larger separation, the behavior of $\Delta v^{n,-}g$
 is dictated by  $\Delta v^{n,-}$.\par
\section{Conclusions and outlook}
\label{sec:conclusions}
We performed interface-resolved direct numerical simulations of turbulent channel flow of suspensions of neutrally-buoyant spherical particles. 
Two flow cases are considered, with same Reynolds number and volume fraction, and particle size varied by a factor of $2$. 
%This resulted in two state-of-the-art direct numerical simulations of this kind. 
The simulations are compared to two single-phase reference simulations: (1) the unladen case and (2) the continuum limit of a Newtonian fluid with a viscosity corresponding to the effective viscosity of the suspension under investigation in a laminar shear flow.\par
As observed in \cite{Costa-et-al-PRL-2016}, the main finite size effect in the zero- and first-order Eulerian statistics originate from the near-wall dynamics, in the near-wall particle layer. 
We show here that the larger the particles are, the further away from the wall the effects of the particle layer are significant. The near-wall inhomogeneity due to the geometrical constraints is felt at distances that scale with the particle size at fixed volume fraction, seemingly independent of the local fluid flow dynamics. This results in a smaller inner-to-outer scale separation for the flow with larger particles, clearly featured in several flow statistics.\par
Particles at the wall with significant particle-to-fluid slip velocity create \textit{hot-spots} of high local wall shear (on average about $4$ times larger than the mean) which contribute significantly to the mean wall-stress. These hot-spots show higher magnitude of the wall stress for small particles, which are, consistently, more localized. Their contribution to the mean wall shear is nearly the same for both particle sizes considered, about $70\%$.
Also interesting to note is the highly increased probability of local shear stresses lower than the mean, and even negative (i.e., instantaneous flow reversal). These hot-spots change considerably the distribution of shear stresses, resulting in a pdf with a wide exponential tail and  rms values,  $\tau_w^{rms}\approx \tau_w$, 
in contrast to what found in canonical single-phase wall-bounded turbulent flows, where the pdf is well fitted by a log-normal distribution and $\tau_w^{rms}\approx 0.4\tau_w$. 
To quantify, the probability of finding a value of the shear stress of about $2$ times the total mean is negligible in the reference single-phase flow, whereas it becomes of the order of $10\%$ in the particle-laden flows.
These findings are used to derive a scaling law for the wall particle-to-fluid apparent slip velocity as a function of the flow governing parameters.\par
Profiles of particle Reynolds number based on the particle slip velocity present values in the bulk of $O(1)$ for the case with large particles, and very close to zero for the suspension with smaller particles. This finite particle Reynolds number for case D20 can explain the significantly larger contribution of particle stresses to the total stresses with respect to the continuum limit reference. This can be attributed to the inertial-shear-thickening mechanism described in \cite{Picano-et-al-PRL-2013}, where finite inertia effectively increases the particles' excluded volume.\par
Finite-size effects in the bulk of the flow are apparent in the second-order statistics, as shown by the profiles of the fluid velocity rms and Reynolds stresses. For these, we have tested the scaling arguments of \cite{Costa-et-al-PRL-2016} and found that
 accounting for the particle-wall layer is a sufficient correction for the case with smaller particles, $D_p/\delta_v\approx 10$ (in fact, the finite size effects are relatively small in the bulk flow and using an effective viscosity only can still provide reasonable estimates).
  The flow with larger particles ($D_p/\delta_v\approx 20$), however,  shows clear deviations from the continuum limit in the bulk of the channel also with the proposed rescaling, despite the fact that the
  correction for the particle-wall layer is sufficient for  the  lower-order statistics.\par
When investigating the particle dynamics, we compute statistics in wall-parallel bins of  small wall-normal extent
to take into account the effects of an inhomogeneous mean flow. Single-point dispersion statistics show that the spanwise particle dispersion coefficient is fairly independent of the wall-normal location, except very close to the wall. The spanwise particle dispersion is attributed to particle-turbulence interactions because of, first, the smaller diffusion coefficient near the wall and second, the larger diffusion coefficients for the smaller particles. The first effect is a
 consequence of the smaller spectrum of turbulence scales capable of disrupting the particle motion, whereas the second to the wider range of scales able to significantly displace the particles.\par
Conversely, the two-point dispersion statistics strongly depend on the wall-normal location and thus on the local shear rate. Higher shear rates induce larger relative velocities which result in faster dispersion. Even though the particles have a finite size, their dispersion statistics at short times show an exponential growth of the absolute displacement with time -- as observed for point-particles with sub-Kolmogorov separation distances. The mechanism is, however, different. In this case short-range inter-particle interactions are likely the cause. Faster pair-dispersion for the larger particles is linked to the larger inter-particle interaction velocity.\par
Finally we investigated the particle collisional dynamics. Larger particles show higher mean values of relative velocity towards contact, consistent with the picture of \pc{viscous-dominated}, shear-induced contacts. This also explains the higher probability of finding pair of larger particles at close distance. Wall-normal variations in collision probability are therefore a consequence of the variations in local shear, and thus the larger particles collide more frequently than the small particles.\par
In this work, we have  explored the advantages of massively parallel simulations, which allow for a  multi-scale, three-dimensional and time-resolved picture of a system with well-defined physical parameters. This and similar studies show that the community have reached a point when simulations of  interface-resolved particle-laden flows are possible. Such simulations, yet computationally expensive, can serve as valuable tool for validation of simpler two-way coupling algorithms and perhaps Eulerian models; we believe this type of simulations will spawn several investigations aiming at better models.
\section{Acknowledgements}
This work was supported by the Portuguese Foundation for Science and Technology under the grant No.\ SFRH/BD/85501/2012, by the European Research Council Grant No. ERC-2013-CoG-616186, TRITOS, by the Swedish Research Council (VR) and by COST Action MP1305: Flowing Matter. We acknowledge computer time provided by SNIC (Swedish National Infrastructure for Computing) and PRACE project 2014112543 for awarding us access to resource CURIE based in France at Genci/CEA.
%%fakesection bibliography
\bibliography{bibfile}

\begin{thebibliography}{60}
\expandafter\ifx\csname natexlab\endcsname\relax\def\natexlab#1{#1}\fi
\def\au#1{#1} \def\ed#1{#1} \def\yr#1{#1}\def\at#1{#1}\def\jt#1{\textit{#1}}
  \def\bt#1{#1}\def\bvol#1{\textbf{#1}} \def\vol#1{#1} \def\pg#1{#1}
  \def\publ#1{#1}\def\arxiv#1{#1}\def\org#1{#1}\def\st#1{\textit{#1}}

\bibitem[Ardekani {\em et~al.\/}(2017)Ardekani, Costa, Breugem, Picano \&
  Brandt]{Ardekani-et-al-JFM-2016}
{\sc \au{Ardekani, M~Niazi}, \au{Costa, Pedro}, \au{Breugem, W-P}, \au{Picano,
  Francesco} \& \au{Brandt, Luca}} \yr{2017}  \at{Drag reduction in turbulent
  channel flow laden with finite-size oblate spheroids}.  \jt{Journal of Fluid
  Mechanics}  \bvol{816},  \pg{43--70}.

\bibitem[Bagnold(1954)]{Bagnold-1954}
{\sc \au{Bagnold, Ralph~A}} \yr{1954} Experiments on a gravity-free dispersion
  of large solid spheres in a newtonian fluid under shear.  \bt{In {\em
  Proceedings of the Royal Society of London A: Mathematical, Physical and
  Engineering Sciences\/}}, ,  \vol{vol. 225},  \pg{pp. 49--63}. The Royal
  Society.

\bibitem[Batchelor(1952)]{Batchelor1952}
{\sc \au{Batchelor, GK}} \yr{1952} The effect of homogeneous turbulence on
  material lines and surfaces.  \bt{In {\em Proceedings of the Royal Society of
  London A: Mathematical, Physical and Engineering Sciences\/}}, ,  \vol{vol.
  213},  \pg{pp. 349--366}. The Royal Society.

\bibitem[Batchelor(1970)]{Batchelor-JFM-1970}
{\sc \au{Batchelor, GK}} \yr{1970}  \at{The stress system in a suspension of
  force-free particles}.  \jt{Journal of fluid mechanics}  \bvol{41}~(03),
  \pg{545--570}.

\bibitem[Biegert {\em et~al.\/}(2017)Biegert, Vowinckel \&
  Meiburg]{Biegert-et-al-JCP-2017}
{\sc \au{Biegert, Edward}, \au{Vowinckel, Bernhard} \& \au{Meiburg, Eckart}}
  \yr{2017}  \at{A collision model for grain-resolving simulations of flows
  over dense, mobile, polydisperse granular sediment beds}.  \jt{Journal of
  Computational Physics}  \bvol{340},  \pg{105--127}.

\bibitem[Blanc {\em et~al.\/}(2013)Blanc, Lemaire, Meunier \&
  Peters]{Blanc-et-al-JoR-2013}
{\sc \au{Blanc, Fr{\'e}d{\'e}ric}, \au{Lemaire, Elisabeth}, \au{Meunier, Alain}
  \& \au{Peters, Fran{\c{c}}ois}} \yr{2013}  \at{Microstructure in sheared
  non-brownian concentrated suspensions}.  \jt{Journal of rheology}
  \bvol{57}~(1),  \pg{273--292}.

\bibitem[Brady \& Bossis(1988)]{Brady-et-al-ARFM-1988}
{\sc \au{Brady, John~F} \& \au{Bossis, Georges}} \yr{1988}  \at{Stokesian
  dynamics}.  \jt{Annual review of fluid mechanics}  \bvol{20},  \pg{111--157}.

\bibitem[Breugem(2012)]{Breugem-JCP-2012}
{\sc \au{Breugem, Wim-Paul}} \yr{2012}  \at{A second-order accurate immersed
  boundary method for fully resolved simulations of particle-laden flows}.
  \jt{Journal of Computational Physics}  \bvol{231}~(13),  \pg{4469--4498}.

\bibitem[Brown \& Jaeger(2014)]{Brown-et-al-RPP-2014}
{\sc \au{Brown, Eric} \& \au{Jaeger, Heinrich~M}} \yr{2014}  \at{Shear
  thickening in concentrated suspensions: phenomenology, mechanisms and
  relations to jamming}.  \jt{Reports on Progress in Physics}  \bvol{77}~(4),
  \pg{046602}.

\bibitem[Chouippe \& Uhlmann(2015)]{Chouippe-and-Uhlmann-PoF-2015}
{\sc \au{Chouippe, Agathe} \& \au{Uhlmann, Markus}} \yr{2015}  \at{Forcing
  homogeneous turbulence in direct numerical simulation of particulate flow
  with interface resolution and gravity}.  \jt{Physics of Fluids
  (1994-present)}  \bvol{27}~(12),  \pg{123301}.

\bibitem[Costa {\em et~al.\/}(2015)Costa, Boersma, Westerweel \&
  Breugem]{Costa-et-al-PRE-2015}
{\sc \au{Costa, Pedro}, \au{Boersma, Bendiks~Jan}, \au{Westerweel, Jerry} \&
  \au{Breugem, Wim-Paul}} \yr{2015}  \at{Collision model for fully resolved
  simulations of flows laden with finite-size particles}.  \jt{Physical Review
  E}  \bvol{92}~(5),  \pg{053012}.

\bibitem[Costa {\em et~al.\/}(2016)Costa, Picano, Brandt \&
  Breugem]{Costa-et-al-PRL-2016}
{\sc \au{Costa, Pedro}, \au{Picano, Francesco}, \au{Brandt, Luca} \&
  \au{Breugem, Wim-Paul}} \yr{2016}  \at{Universal scaling laws for dense
  particle suspensions in turbulent wall-bounded flows}.  \jt{Phys. Rev. Lett.}
   \bvol{117},  \pg{134501}.

\bibitem[Crowe {\em et~al.\/}(1977)Crowe, Sharma \&
  Stock]{Crowe-et-al-JFE-1977}
{\sc \au{Crowe, Clayton~T}, \au{Sharma, M~Pt} \& \au{Stock, David~E}} \yr{1977}
   \at{The particle-source-in cell (psi-cell) model for gas-droplet flows}.
  \jt{Journal of Fluids Engineering}  \bvol{99}~(2),  \pg{325--332}.

\bibitem[Dance \& Maxey(2003)]{Dance-and-Maxey-JCP-2003}
{\sc \au{Dance, SL} \& \au{Maxey, MR}} \yr{2003}  \at{Incorporation of
  lubrication effects into the force-coupling method for particulate two-phase
  flow}.  \jt{Journal of computational Physics}  \bvol{189}~(1),
  \pg{212--238}.

\bibitem[Eaton \& Fessler(1994)]{Eaton-and-Fessler-IJMF-1994}
{\sc \au{Eaton, John~K} \& \au{Fessler, JR}} \yr{1994}  \at{Preferential
  concentration of particles by turbulence}.  \jt{International Journal of
  Multiphase Flow}  \bvol{20},  \pg{169--209}.

\bibitem[Einstein(1905)]{Einstein1905}
{\sc \au{Einstein, Albert}} \yr{1905}  \at{{\"U}ber die von der
  molekularkinetischen theorie der w{\"a}rme geforderte bewegung von in
  ruhenden fl{\"u}ssigkeiten suspendierten teilchen}.  \jt{Annalen der physik}
  \bvol{322}~(8),  \pg{549--560}.

\bibitem[Eshghinejadfard {\em et~al.\/}(2017)Eshghinejadfard, Hosseini \&
  Th{\'e}venin]{Eshghinejadfard-AIP-2017}
{\sc \au{Eshghinejadfard, Amir}, \au{Hosseini, Seyed~Ali} \& \au{Th{\'e}venin,
  Dominique}} \yr{2017}  \at{Fully-resolved prolate spheroids in turbulent
  channel flows: A lattice boltzmann study}.  \jt{AIP Advances}  \bvol{7}~(9),
  \pg{095007}.

\bibitem[Fessler {\em et~al.\/}(1994)Fessler, Kulick \&
  Eaton]{Fessler-et-al-Pof-1994}
{\sc \au{Fessler, John~R}, \au{Kulick, Jonathan~D} \& \au{Eaton, John~K}}
  \yr{1994}  \at{Preferential concentration of heavy particles in a turbulent
  channel flow}.  \jt{Physics of Fluids (1994-present)}  \bvol{6}~(11),
  \pg{3742--3749}.

\bibitem[Foerster {\em et~al.\/}(1994)Foerster, Louge, Chang \&
  Allia]{Foerster-et-al-PoF-1994}
{\sc \au{Foerster, Samuel~F}, \au{Louge, Michel~Y}, \au{Chang, Hongder} \&
  \au{Allia, Khedidja}} \yr{1994}  \at{Measurements of the collision properties
  of small spheres}.  \jt{Physics of Fluids}  \bvol{6}~(3),  \pg{1108--1115}.

\bibitem[Fornari {\em et~al.\/}(2016{\natexlab{{\em a\/}}})Fornari, Formenti,
  Picano \& Brandt]{Fornari-et-al-PoF-2016}
{\sc \au{Fornari, W.}, \au{Formenti, A.}, \au{Picano, F.} \& \au{Brandt, L.}}
  \yr{2016{\natexlab{{\em a\/}}}}  \at{The effect of particle density in
  turbulent channel flow laden with finite size particles in semi-dilute
  conditions}.  \jt{Physics of Fluids}  \bvol{28}~(3).

\bibitem[Fornari {\em et~al.\/}(2016{\natexlab{{\em b\/}}})Fornari, Picano \&
  Brandt]{Fornari-et-al-JFM-2016}
{\sc \au{Fornari, Walter}, \au{Picano, Francesco} \& \au{Brandt, Luca}}
  \yr{2016{\natexlab{{\em b\/}}}}  \at{Sedimentation of finite-size spheres in
  quiescent and turbulent environments}.  \jt{Journal of Fluid Mechanics}
  \bvol{788},  \pg{640--669}.

\bibitem[Fornari {\em et~al.\/}(2018)Fornari, Picano \&
  Brandt]{Fornari-Poly-New}
{\sc \au{Fornari, Walter}, \au{Picano, Francesco} \& \au{Brandt, Luca}}
  \yr{2018}  \at{The effect of polydispersity in a turbulent channel flow laden
  with finite-size particles}.  \jt{European Journal of Mechanics - B/Fluids}
  \bvol{67}~(Supplement C),  \pg{54 -- 64}.

\bibitem[Guazzelli \& Morris(2011)]{Guazzelli-and-Morris-2011}
{\sc \au{Guazzelli, Elisabeth} \& \au{Morris, Jeffrey~F}} \yr{2011} {\em A
  physical introduction to suspension dynamics\/}, ,  \vol{vol.~45}.
  \publ{Cambridge University Press}.

\bibitem[Hinze(1975)]{Hinze-1975}
{\sc \au{Hinze, J.O.}} \yr{1975} {\em Turbulence\/},  \st{McGraw-Hill classic
  textbook reissue series},  \vol{vol. 218}.  \publ{McGraw-Hill}.

\bibitem[Hunt {\em et~al.\/}(2002)Hunt, Zenit, Campbell \&
  Brennen]{Hunt-et-al-JFM-2002}
{\sc \au{Hunt, ML}, \au{Zenit, R}, \au{Campbell, CS} \& \au{Brennen, CE}}
  \yr{2002}  \at{Revisiting the 1954 suspension experiments of ra bagnold}.
  \jt{Journal of Fluid Mechanics}  \bvol{452},  \pg{1--24}.

\bibitem[Joseph {\em et~al.\/}(2001)Joseph, Zenit, Hunt \&
  Rosenwinkel]{Joseph-et-al-JFM-2001}
{\sc \au{Joseph, GG}, \au{Zenit, R}, \au{Hunt, ML} \& \au{Rosenwinkel, AM}}
  \yr{2001}  \at{Particle--wall collisions in a viscous fluid}.  \jt{Journal of
  Fluid Mechanics}  \bvol{433},  \pg{329--346}.

\bibitem[Kidanemariam \& Uhlmann(2014)]{Kidanemariam-and-Uhlmann-JFM-2014}
{\sc \au{Kidanemariam, Aman~G} \& \au{Uhlmann, Markus}} \yr{2014}  \at{Direct
  numerical simulation of pattern formation in subaqueous sediment}.
  \jt{Journal of Fluid Mechanics}  \bvol{750},  \pg{R2}.

\bibitem[Kim {\em et~al.\/}(1987)Kim, Moin \& Moser]{Kim-et-al-JFM-1987}
{\sc \au{Kim, John}, \au{Moin, Parviz} \& \au{Moser, Robert}} \yr{1987}
  \at{Turbulence statistics in fully developed channel flow at low reynolds
  number}.  \jt{Journal of fluid mechanics}  \bvol{177},  \pg{133--166}.

\bibitem[Lashgari {\em et~al.\/}(2014)Lashgari, Picano, Breugem \&
  Brandt]{Lashgari-et-al-PRL-2014}
{\sc \au{Lashgari, Iman}, \au{Picano, Francesco}, \au{Breugem, Wim-Paul} \&
  \au{Brandt, Luca}} \yr{2014}  \at{Laminar, turbulent, and inertial
  shear-thickening regimes in channel flow of neutrally buoyant particle
  suspensions}.  \jt{Phys. Rev. Lett.}  \bvol{113},  \pg{254502}.

\bibitem[Lashgari {\em et~al.\/}(2016)Lashgari, Picano, Breugem \&
  Brandt]{Lashgari-et-al-IJMF-2016}
{\sc \au{Lashgari, Iman}, \au{Picano, Francesco}, \au{Breugem, Wim~Paul} \&
  \au{Brandt, Luca}} \yr{2016}  \at{Channel flow of rigid sphere suspensions:
  Particle dynamics in the inertial regime}.  \jt{International Journal of
  Multiphase Flow}  \bvol{78},  \pg{12--24}.

\bibitem[Lashgari {\em et~al.\/}(2017)Lashgari, Picano, Costa, Breugem \&
  Brandt]{Lashgari-et-al-JFM-2016}
{\sc \au{Lashgari, Iman}, \au{Picano, Francesco}, \au{Costa, Pedro},
  \au{Breugem, Wim-Paul} \& \au{Brandt, Luca}} \yr{2017}  \at{Turbulent channel
  flow of a dense binary mixture of rigid particles}.  \jt{Journal of Fluid
  Mechanics}  \bvol{818},  \pg{623--645}.

\bibitem[Legendre {\em et~al.\/}(2006)Legendre, Zenit, Daniel \&
  Guiraud]{Legendre-et-al-CES-2006}
{\sc \au{Legendre, Dominique}, \au{Zenit, Roberto}, \au{Daniel, Claude} \&
  \au{Guiraud, Pascal}} \yr{2006}  \at{A note on the modelling of the bouncing
  of spherical drops or solid spheres on a wall in viscous fluid}.
  \jt{Chemical engineering science}  \bvol{61}~(11),  \pg{3543--3549}.

\bibitem[Leighton \& Acrivos(1987)]{Leighton-and-Acrivos-JFM-1987}
{\sc \au{Leighton, David} \& \au{Acrivos, Andreas}} \yr{1987}  \at{The
  shear-induced migration of particles in concentrated suspensions}.
  \jt{Journal of Fluid Mechanics}  \bvol{181},  \pg{415--439}.

\bibitem[Loisel {\em et~al.\/}(2013)Loisel, Abbas, Masbernat \&
  Climent]{Loisel-et-al-PoF-2013}
{\sc \au{Loisel, Vincent}, \au{Abbas, Micheline}, \au{Masbernat, Olivier} \&
  \au{Climent, Eric}} \yr{2013}  \at{The effect of neutrally buoyant
  finite-size particles on channel flows in the laminar-turbulent transition
  regime}.  \jt{Physics of Fluids}  \bvol{25}~(12),  \pg{123304}.

\bibitem[Lucci {\em et~al.\/}(2010)Lucci, Ferrante \&
  Elghobashi]{Lucci-et-al-JFM-2010}
{\sc \au{Lucci, Francesco}, \au{Ferrante, Antonino} \& \au{Elghobashi, Said}}
  \yr{2010}  \at{Modulation of isotropic turbulence by particles of taylor
  length-scale size}.  \jt{Journal of Fluid Mechanics}  \bvol{650},
  \pg{5--55}.

\bibitem[Marchioro {\em et~al.\/}(1999)Marchioro, Tanksley \&
  Prosperetti]{Marchioro-et-al-IJMF-1999}
{\sc \au{Marchioro, M}, \au{Tanksley, M} \& \au{Prosperetti, A}} \yr{1999}
  \at{Mixture pressure and stress in disperse two-phase flow}.
  \jt{International journal of multiphase flow}  \bvol{25}~(6),
  \pg{1395--1429}.

\bibitem[Matas {\em et~al.\/}(2003)Matas, Morris \&
  Guazzelli]{Matas-et-al-PRL-2003}
{\sc \au{Matas, J-P}, \au{Morris, Jeffrey~F} \& \au{Guazzelli, E}} \yr{2003}
  \at{Transition to turbulence in particulate pipe flow}.  \jt{Physical review
  letters}  \bvol{90}~(1),  \pg{014501}.

\bibitem[Maxey \& Riley(1983)]{Maxey-and-Riley-PoF-1983}
{\sc \au{Maxey, Martin~R} \& \au{Riley, James~J}} \yr{1983}  \at{Equation of
  motion for a small rigid sphere in a nonuniform flow}.  \jt{Physics of Fluids
  (1958-1988)}  \bvol{26}~(4),  \pg{883--889}.

\bibitem[Morris(2009)]{Morris-RA-2009}
{\sc \au{Morris, Jeffrey~F}} \yr{2009}  \at{A review of microstructure in
  concentrated suspensions and its implications for rheology and bulk flow}.
  \jt{Rheologica acta}  \bvol{48}~(8),  \pg{909--923}.

\bibitem[\"Orl\"u \& Schlatter(2011)]{Orlu-and-Schlatter-PoF-2011}
{\sc \au{\"Orl\"u, Ramis} \& \au{Schlatter, Philipp}} \yr{2011}  \at{On the
  fluctuating wall-shear stress in zero pressure-gradient turbulent boundary
  layer flows}.  \jt{Physics of fluids}  \bvol{23}~(2),  \pg{021704}.

\bibitem[Picano {\em et~al.\/}(2015)Picano, Breugem \&
  Brandt]{Picano-et-al-JFM-2015}
{\sc \au{Picano, Francesco}, \au{Breugem, Wim-Paul} \& \au{Brandt, Luca}}
  \yr{2015}  \at{Turbulent channel flow of dense suspensions of neutrally
  buoyant spheres}.  \jt{Journal of Fluid Mechanics}  \bvol{764},
  \pg{463--487}.

\bibitem[Picano {\em et~al.\/}(2013)Picano, Breugem, Mitra \&
  Brandt]{Picano-et-al-PRL-2013}
{\sc \au{Picano, Francesco}, \au{Breugem, Wim-Paul}, \au{Mitra, Dhrubaditya} \&
  \au{Brandt, Luca}} \yr{2013}  \at{Shear thickening in non-brownian
  suspensions: An excluded volume effect}.  \jt{Phys. Rev. Lett.}  \bvol{111},
  \pg{098302}.

\bibitem[Pope(2001)]{Pope-2001}
{\sc \au{Pope, Stephen~B}} \yr{2001} Turbulent flows.

\bibitem[Prosperetti(2015)]{Prosperetti-JFM-2015}
{\sc \au{Prosperetti, Andrea}} \yr{2015}  \at{Life and death by boundary
  conditions}.  \jt{Journal of fluid mechanics}  \bvol{768},  \pg{1--4}.

\bibitem[Reeks(1977)]{Reeks-JFM-1977}
{\sc \au{Reeks, MW}} \yr{1977}  \at{On the dispersion of small particles
  suspended in an isotropic turbulent fluid}.  \jt{Journal of fluid mechanics}
  \bvol{83}~(03),  \pg{529--546}.

\bibitem[Roma {\em et~al.\/}(1999)Roma, Peskin \& Berger]{Roma-et-al-JCP-1999}
{\sc \au{Roma, Alexandre~M}, \au{Peskin, Charles~S} \& \au{Berger, Marsha~J}}
  \yr{1999}  \at{An adaptive version of the immersed boundary method}.
  \jt{Journal of Computational Physics}  \bvol{153}~(2),  \pg{509--534}.

\bibitem[Salazar \& Collins(2009)]{Salazar2009}
{\sc \au{Salazar, Juan~PLC} \& \au{Collins, Lance~R}} \yr{2009}
  \at{Two-particle dispersion in isotropic turbulent flows}.  \jt{Annual review
  of fluid mechanics}  \bvol{41},  \pg{405--432}.

\bibitem[Sardina {\em et~al.\/}(2012)Sardina, Schlatter, Brandt, Picano \&
  Casciola]{Sardina-et-al-JFM-2012}
{\sc \au{Sardina, G}, \au{Schlatter, Philipp}, \au{Brandt, Luca}, \au{Picano,
  F} \& \au{Casciola, CM}} \yr{2012}  \at{Wall accumulation and spatial
  localization in particle-laden wall flows}.  \jt{Journal of Fluid Mechanics}
  \bvol{699},  \pg{50--78}.

\bibitem[Sierou \& Brady(2004)]{Sierou-and-Brady-JFM-2004}
{\sc \au{Sierou, Asimina} \& \au{Brady, John~F}} \yr{2004}  \at{Shear-induced
  self-diffusion in non-colloidal suspensions}.  \jt{Journal of fluid
  mechanics}  \bvol{506}~(1),  \pg{285--314}.

\bibitem[Soldati \& Marchioli(2009)]{Soldati-and-Marchioli-IJMF-2009}
{\sc \au{Soldati, Alfredo} \& \au{Marchioli, Cristian}} \yr{2009}  \at{Physics
  and modelling of turbulent particle deposition and entrainment: Review of a
  systematic study}.  \jt{International Journal of Multiphase Flow}
  \bvol{35}~(9),  \pg{827--839}.

\bibitem[Stickel \& Powell(2005)]{Stickel-and-Powell-ARFM-2005}
{\sc \au{Stickel, Jonathan~J} \& \au{Powell, Robert~L}} \yr{2005}  \at{Fluid
  mechanics and rheology of dense suspensions}.  \jt{Annu. Rev. Fluid Mech.}
  \bvol{37},  \pg{129--149}.

\bibitem[Ten~Cate {\em et~al.\/}(2004)Ten~Cate, Derksen, Portela \& Van
  Den~Akker]{TenCate-et-al-JFM-2004}
{\sc \au{Ten~Cate, Andreas}, \au{Derksen, Jos~J}, \au{Portela, Luis~M} \&
  \au{Van Den~Akker, Harry~EA}} \yr{2004}  \at{Fully resolved simulations of
  colliding monodisperse spheres in forced isotropic turbulence}.  \jt{J. Fluid
  Mech}  \bvol{519},  \pg{233--271}.

\bibitem[Uhlmann(2005)]{Uhlmann-JCP-2005}
{\sc \au{Uhlmann, Markus}} \yr{2005}  \at{An immersed boundary method with
  direct forcing for the simulation of particulate flows}.  \jt{Journal of
  Computational Physics}  \bvol{209}~(2),  \pg{448--476}.

\bibitem[Uhlmann(2008)]{Uhlmann-PoF-2008}
{\sc \au{Uhlmann, Markus}} \yr{2008}  \at{Interface-resolved direct numerical
  simulation of vertical particulate channel flow in the turbulent regime}.
  \jt{Physics of Fluids (1994-present)}  \bvol{20}~(5),  \pg{053305}.

\bibitem[Vowinckel {\em et~al.\/}(2014)Vowinckel, Kempe \&
  Fr{\"o}hlich]{Vowinckel-et-al-AWR-2014}
{\sc \au{Vowinckel, Bernhard}, \au{Kempe, Tobias} \& \au{Fr{\"o}hlich, Jochen}}
  \yr{2014}  \at{Fluid--particle interaction in turbulent open channel flow
  with fully-resolved mobile beds}.  \jt{Advances in Water Resources}
  \bvol{72},  \pg{32--44}.

\bibitem[Vreman(2016)]{Vreman-JFM-2016}
{\sc \au{Vreman, A.~W.}} \yr{2016}  \at{Particle-resolved direct numerical
  simulation of homogeneous isotropic turbulence modified by small fixed
  spheres}.  \jt{Journal of Fluid Mechanics}  \bvol{796},  \pg{40--85}.

\bibitem[Wang {\em et~al.\/}(2017)Wang, Abbas \& Climent]{Wang-et-al-PRF-2017}
{\sc \au{Wang, Guiquan}, \au{Abbas, Micheline} \& \au{Climent, Eric}} \yr{2017}
   \at{Modulation of large-scale structures by neutrally buoyant and inertial
  finite-size particles in turbulent couette flow}.  \jt{Physical Review
  Fluids}  \bvol{2}~(8),  \pg{084302}.

\bibitem[Wang {\em et~al.\/}(2016)Wang, Peng, Guo \& Yu]{Wang-et-al-JFE-2016}
{\sc \au{Wang, Lian-Ping}, \au{Peng, Cheng}, \au{Guo, Zhaoli} \& \au{Yu,
  Zhaosheng}} \yr{2016}  \at{Flow modulation by finite-size neutrally buoyant
  particles in a turbulent channel flow}.  \jt{Journal of Fluids Engineering}
  \bvol{138}~(4),  \pg{041306}.

\bibitem[Yang \& Hunt(2006)]{Yang-and-Hunt-PoF-2006}
{\sc \au{Yang, F-L} \& \au{Hunt, ML}} \yr{2006}  \at{Dynamics of
  particle-particle collisions in a viscous liquid}.  \jt{Physics of Fluids}
  \bvol{18}~(12),  \pg{121506}.

\bibitem[Yu {\em et~al.\/}(2016)Yu, Vinkovic \& Buffat]{Yu-et-al-JoT-2016}
{\sc \au{Yu, W}, \au{Vinkovic, I} \& \au{Buffat, M}} \yr{2016}  \at{Finite-size
  particles in turbulent channel flow: quadrant analysis and acceleration
  statistics}.  \jt{Journal of Turbulence}  \bvol{17}~(11),  \pg{1048--1071}.

\end{thebibliography}
\end{document}